\documentclass{iopart}

\usepackage[english]{babel}
\usepackage[dvips]{graphicx}
\usepackage{amssymb}
\usepackage{amsthm}
\usepackage{bm}
\usepackage{cite}
\usepackage{color}
\usepackage{iopams}
\usepackage{latexsym}
\usepackage{times}
\usepackage{stmaryrd}
\usepackage{url}

\usepackage{hyperref}
\hypersetup{
  colorlinks=true,        
  linkcolor=blue,         
  citecolor=cyan,         
}

\newcommand{\eqref}[1]{(\ref{#1})}

\newcommand{\bF}{\boldsymbol{F}}
\newcommand{\bS}{\boldsymbol{S}}
\newcommand{\bu}{\boldsymbol{u}}
\newcommand{\bx}{\boldsymbol{x}}

\newcommand{\dd}{{d}}

\newcommand{\cf}{\textit{cf.}~}
\newcommand{\ie}{\textit{i.e.}~}
\newcommand{\eg}{\textit{e.g.}~}

\begin{document}

\title{High-Order Fully General-Relativistic Hydrodynamics: new
  Approaches and Tests}

\author{David Radice$^{1,2}$, Luciano Rezzolla$^{3,2}$ and Filippo
Galeazzi$^{2,4}$}

\address{$^1$ Theoretical Astrophysics, California Institute of
Technology, 1200 E California Blvd, Pasadena, California 91125, USA}

\address{$^2$ Max-Planck-Institut f\"ur Gravitationsphysik, Albert
Einstein Institut, Am M\"uhlenberg 1, 14476 Potsdam, Germany}

\address{$^3$ Institut f\"ur Theoretische Physik, Max-von-Laue-Str. 1,
  60438 Frankfurt am Main, Germany}

\address{$^4$ Departamento de Astronom\'{\i}a y Astrof\'{\i}sica,
Universitat de Val\`encia, Dr. Moliner 50, 46100, Burjassot (Val\`encia),
Spain}

\begin{abstract}
We present a new approach for achieving high-order convergence in fully
general-relativistic hydrodynamic simulations. The approach is
implemented in \texttt{WhiskyTHC}, a new code that makes use of
state-of-the-art numerical schemes and was key in achieving, for the
first time, higher than second-order convergence in the calculation of
the gravitational radiation from inspiraling binary neutron stars
\cite{Radice2013b}. Here, we give a detailed description of the
algorithms employed and present results obtained for a series of
classical tests involving isolated neutron stars. In addition, using the
gravitational-wave emission from the late inspiral and merger of binary
neutron stars, we make a detailed comparison between the results obtained
with the new code and those obtained when using standard second-order
schemes commonly employed for matter simulations in numerical
relativity. We find that even at moderate resolutions and for
binaries with large compactness, the phase accuracy is improved by a
factor $50$ or more.
\end{abstract}

\pacs{
04.25.Dm, 
04.30.Db, 
95.30.Lz, 
95.30.Sf 
}

\section{Introduction}

The accurate modeling of gravitational waves from the late-inspiral and
merger of binary neutron stars can only be achieved within the framework
of numerical relativity and exploiting the tools of computational
relativistic hydrodynamics. Assuming the simplest physical scenarios, \ie
irrotational binaries in quasi-circular orbits, and idealized equations
of state, \ie polytropic of Gamma-law, very accurate numerical-relativity
waveforms for binary neutron stars are nowadays available, \eg
\cite{Baiotti:2010, Baiotti2011, Bernuzzi2012, Hotokezaka2013b,
  Radice2013b}. However, obtaining good-quality waveforms to fully cover
the large parameter space of possible binary-neutron-star configurations,
equations of state and compactnesses, seems to be out of the reach for
current-generation codes. The main reason behind this difficulty is to be
found in the small convergence order, \ie $\lesssim 2$, typical of
general-relativistic hydrodynamics codes, and which has a number of
undesired consequences. Among these, the fact that obtaining
high-accuracy waveforms from low-order codes requires very high spatial
resolutions (and hence very high computational costs), or the fact that
the analysis of the waveforms is spoiled by the large phase uncertainties
typical of these simulations. Both of these difficulties can be resolved
in part by employing new, state-of-the-art schemes that are able to go
over the second-order of convergence typical of general-relativistic
hydrodynamics codes. Using these techniques, we were recently able to
achieve, for the first time, higher than second-order convergence in both
the phase and amplitude of the gravitational-wave radiation from binary
neutron stars \cite{Radice2013b}. These new advances could potentially
enable a more systematic study of the gravitational radiation from binary
neutron stars, similarly to what has been done for the case of binary
black-holes, \eg \cite{Aylott:2009ya, Hinder2013}.

The goal of this paper is to give a full description of the methods we
employed in \cite{Radice2013b} and to describe in detail our new
high-order, high-resolution shock-capturing, finite-differencing code:
\texttt{WhiskyTHC}, which represents the extension to general relativity
of the \texttt{THC} code presented in \cite{Radice2012a}. When compared
with other high-order relativistic hydrodynamics codes, such as
\texttt{wham} \cite{Tchekhovskoy2007} and \texttt{ECHO}
\cite{DelZanna2007, Bucciantini2011}, this is the first
higher-than-second-order code that works in full general relativity, \ie
with dynamical spacetime, and in three spatial dimensions.

First, we demonstrate the capabilities of the new code in a series of
tests involving the evolution of isolated neutron stars. More
specifically, we show that the code is able to yield long-term and
accurate evolutions of stable and unstable stars. We measure the accuracy
of the code for the case of an unstable star collapsing to a black hole
and show that we are able to achieve third-order convergence.

Second, we consider the performance of the code in calculating the
inspiral and merger of binary neutron stars in quasi-circular orbits. We
show higher than second-order convergence for the phase and the amplitude
of the gravitational waves produced in this process. When compared with
the performance of our previous \texttt{Whisky} code, which adopts the
standard second-order schemes commonly employed for matter simulations in
numerical relativity, we can show that the new code is able to yield a
decrease in the phase error of a factor $\sim 50$ for simulations with
the same resolution and with similar computational costs.

The rest of this paper is organized as follows. In Section
\ref{sec:grthc.hydro} we quickly recall the equations of
general-relativistic hydrodynamics and the CCZ4 formalism used to evolve
the spacetime for the tests presented here. In Section
\ref{sec:grthc.code} we give a quick summary of the numerical methods
that we employed, as well as a detailed description of the treatment of
the fluid--vacuum interfaces, which was one of the main challenges in the
application of higher-order numerical schemes to binary-neutron-stars
simulations. In Section \ref{sec:grthc.sns} we present the results
obtained with our code in a series of representative tests involving the
evolution of isolated neutron stars, with particular focus on the
properties of the different vacuum treatments implemented in the code. In
Section \ref{sec:grthc.bns} we present results obtained in the case of
binary neutron stars and, finally, we dedicate Section
\ref{sec:grthc.conclusions} to the discussion of our results and to the
conclusions.

We use a spacetime signature $(-,+,+,+)$, with Greek indices running from
0 to 3 and Latin indices from 1 to 3. We also employ the standard
convention for the summation over repeated indices. Unless otherwise
stated, all quantities are expressed in a system of units in which
$c=G=1$.

\section{General-Relativistic Hydrodynamics}
\label{sec:grthc.hydro}

In this work we adopt the usual 3+1 formalism, \eg \cite{MTW1973}, to
decompose spacetime into space-like hypersurfaces with normal $n^\mu =
(1/\alpha, -\beta^i/\alpha)$, where $\alpha$ is the lapse function and
$\beta^i$ is the shift vector. Within this formalism the spacetime metric
$g_{\mu\nu}$ is split as
\begin{equation}
  \dd s^2 = g_{\mu\nu} \dd x^\mu \dd x^\nu \nonumber 
          = - (\alpha^2 - \beta_i \beta^i) \dd t +2 \beta_i \dd x^i \dd
          t + \gamma_{ij} \dd x^i \dd x^j\,,
\end{equation}
where $\gamma_{ij}$ is the spatial three-metric metric, which, together
with the extrinsic curvature $K_{ij} = - \frac{1}{2} \mathcal{L}_n
\gamma_{ij}$, $\mathcal{L}_n$ being the Lie derivative along $n^\mu$,
fully determines the geometry of each leaf of the foliation. Furthermore
we will indicate with $R_{ij}$ and $R$ the Ricci tensor and scalar
associated with the Levi-Civita connection on the spacelike hypersurfaces
of the foliation, respectively.

The matter content of the spacetime is described through its
energy-momentum tensor $T_{\mu\nu}$, which, within the 3+1 split of
spacetime, can be decomposed in its time, spatial and mixed components as
(see, \eg, \cite{Rezzolla_book:2013})
\begin{equation}
  E      = n^\mu n^\nu T_{\mu\nu}\,,\qquad
  S^i    = \gamma^{i\mu} n^\nu T_{\mu\nu}\,,\qquad
  S^{ij} = \gamma^{i\mu} \gamma^{j\nu} T_{\mu\nu}\,.
\end{equation}
Finally, we will use the convention of raising and lowering indices of
spatial tensors with the (spatial) three-metric, unless otherwise stated.

\subsection{Spacetime Evolution Equations}

For the solution of the Einstein equations we adopt the covariant and
conformal formulation of the Z4 equations (CCZ4). We recall that the Z4
formulation can be obtained from the covariant Lagrangian
\begin{equation}\label{eq:grhydro.Z4.lagrangian}
  \Lambda = g^{\mu\nu} ( R_{\mu\nu} + 2 \nabla_\mu Z_\nu ) \sqrt{-g} +
  \Lambda_m\,,
\end{equation}
by means of a Palatini-type variational principle~\cite{Bona:2010is},
where $g$ is the metric determinant. The variational principle yields
the field equations
\begin{equation}\label{eq:grhydro.Z4.fieldeq}
  R_{\mu\nu} + \nabla_\mu Z_\nu + \nabla_\nu Z_\mu =
    8 \pi \left( T_{\mu\nu} - \frac{1}{2} T g_{\mu\nu} \right)\,,
\end{equation}
as well as a set of constraints fixing the connection
\begin{equation}\label{eq:grhydro.Z4.connection}
  \nabla_\mu g^{\mu\nu} = 0\,,
\end{equation}
and the algebraic constraint
\begin{equation}\label{eq:grhydro.Z4.constraint}
  Z_\mu = 0\,.
\end{equation}
If Eq. \eqref{eq:grhydro.Z4.constraint} is satisfied then
Eqs. \eqref{eq:grhydro.Z4.fieldeq} and \eqref{eq:grhydro.Z4.connection}
reduce to the standard Einstein field equations. Otherwise $Z_\mu$ gives
a measure of the deviation of the solution from the one of the original
Einstein equations. In addition we point out that it is possible to show
that the condition that the first derivatives of $Z_\mu$ vanish amounts
to imposing the ADM momentum and Hamiltonian constraints
\cite{Bona-and-Palenzuela-Luque-2005:numrel-book}.

The key idea of the Z4 formalism is to develop a set of evolution
equations starting from the Lagrangian \eqref{eq:grhydro.Z4.lagrangian},
\emph{without explicitly enforcing \eqref{eq:grhydro.Z4.constraint}}, \ie
treating $Z_\mu$ as a new independent variable. The resulting set of
equations is then strongly hyperbolic, \ie free from the zero-speed modes
of the original ADM system, and the solution of the Einstein equations is
obtained exploiting the fact that the Z4 evolution system preserves the
constraint \eqref{eq:grhydro.Z4.constraint}, \ie $\partial_t (Z_\mu) =
0$. In particular, if the initial-data is constraint satisfying, the Z4
evolution recovers the solution of the Einstein equations, even though
$Z_\mu$ is an evolved variable. In practice, however, small numerical
errors introduce constraint violation during the evolution, for this
reason the Z4 system is usually modified, with the addition of terms that
cancel out in the case in which the constraints are satisfied, to ensure
that eventual constraint violations are propagated away and damped
exponentially \cite{Gundlach2005:constraint-damping}.

The version of Z4 that we employ was recently introduced by
\cite{Alic:2011a} and is based on a conformal decomposition of the
original Z4 system and aims at exploiting well-tested gauge conditions
together with the constraint propagation and damping properties of the
original Z4 formulation. The CCZ4 system then reads \cite{Alic:2011a}
\numparts
\begin{eqnarray}
\fl
\partial_t\tilde\gamma_{ij} = - 2\alpha \tilde A^{^{\rm TF}}_{ij} +
2\tilde\gamma_{k(i}\partial_{j)}~\beta^k -
\frac{2}{3}\tilde\gamma_{ij}\partial_k~\beta^k +\beta^k \partial_k
\tilde\gamma_{ij} \,,  \label{gamma_eq}\\
\fl
\partial_t \tilde A_{ij} = \phi^2  \left[-\nabla_i \nabla_j \alpha +
\alpha \left(R_{ij} + \nabla_i Z_j + \nabla_j Z_i - 8 \pi
S_{ij}\right)\right]^{\rm TF} + \alpha \tilde A_{ij}\left(K-
2\Theta\right) - \nonumber \\
2\alpha \tilde A_{il}\tilde A^l_j A_{ij} \partial_k \beta^k + \beta^k
\partial_k \tilde A_{ij}  \,, \label{A_eq} \\
\fl
\partial_t\phi = \frac{1}{3} \alpha \phi K - \frac{1}{3} \phi
\partial_k \beta^k + \beta^k \partial_k \phi \,, \\
\fl
\partial_t K = - \nabla^i \nabla_i \alpha + \alpha \left(R + 2
\nabla_iZ^i + K^2 -2 \Theta K \right) + \nonumber \\ 
\beta^j \partial_j K - 3 \alpha \kappa_1 \left(1 + \kappa_2\right) \Theta
+ 4 \pi \alpha \left(S - 3 E\right) \,, \label{K_eq}\\
\fl
\partial_t \Theta = \frac{1}{2} \alpha \left(R + 2 \nabla_i Z^i - \tilde
A_{ij} \tilde A^{ij} + \frac{2}{3} K^2 - 2 \Theta K \right) -\nonumber \\ 
Z^i \partial_i \alpha + \beta^k \partial_k \Theta - \alpha \kappa_1
\left(2 + \kappa_2\right) \Theta - 8\pi \alpha\,E\,, \\
\fl
\partial_t \hat\Gamma^i = 2\alpha \left(\tilde\Gamma^i_{jk} \tilde
A^{jk} - 3 \tilde A^{ij} \frac{\partial_j \phi}{\phi} - \frac{2}{3}
\tilde\gamma^{ij} \partial_j K \right) +
2\tilde\gamma^{ki}\left(\alpha \partial_k \Theta - \Theta \partial_k
\alpha -\frac{2}{3} \alpha K Z_k\right) - \nonumber \\  
2\tilde A^{ij} \partial_j \alpha \nonumber + \tilde\gamma^{kl} \partial_k
\partial_l \beta^i + \frac{1}{3}\tilde\gamma^{ik}\partial_k\partial_l
\beta^l + \frac{2}{3} \tilde\Gamma^i \partial_k \beta^k -\nonumber \\ 
\tilde\Gamma^k \partial_k \beta^i + 2 \kappa_3 \left(\frac{2}{3}
\tilde\gamma^{ij} Z_j \partial_k \beta^k - \tilde\gamma^{jk} Z_j
\partial_k \beta^i \right) + \nonumber \\
\beta^k \partial_k \hat\Gamma^i - 2 \alpha \kappa_1 \tilde \gamma^{ij}
Z_j - 16 \pi \alpha \tilde\gamma^{ij} S_{j}\,, \label{Gamma_eq}
\end{eqnarray}
\endnumparts
where $\Theta \equiv n_{\mu} Z^{\mu} = \alpha Z^0$, $S = \gamma^{ij}
S_{ij}$, ${\tilde{\gamma}}_{ij} = \phi^2 \gamma_{ij} $ is the conformal
metric with unit determinant $\phi = ({\rm det} (\gamma_{ij}))^{-1/6}$,
while the extrinsic curvature $K_{ij}$ is split into its trace $K
\equiv K_{ij} \gamma^{ij}$ and its trace-free components
\begin{equation}
\label{tracelessK}
{\tilde{A}}_{ij} = \phi^2\;(K_{ij}- \frac{1}{3} K \gamma_{ij})\,.
\end{equation}

The three-dimensional Ricci tensor $R_{ij}$ is split into a part
containing conformal terms $\tilde R^{\phi}_{ij}$ and another one
containing space derivatives of the conformal metric $\tilde R_{ij}$,
defined as
\begin{eqnarray}
\fl
\qquad \tilde R_{ij} = -\frac{1}{2} \tilde \gamma^{lm} \partial_l \partial_m
\tilde \gamma_{ij} + \tilde \gamma_{k(i} \partial_{j)} \tilde \Gamma^k +
\tilde \Gamma^k \tilde \Gamma_{(ij)k} +
\tilde \gamma^{lm} \left[2 {\tilde \Gamma^k}_{l(i} \tilde \Gamma_{j)km} +
{\tilde \Gamma^k}_{im} \tilde \Gamma_{kj\,l}\right]\,, \\
\fl
\qquad \tilde R^{\phi}_{ij} = \frac{1}{\phi^2}\left[\phi \left(\tilde \nabla_i
\tilde \nabla_j \phi + \tilde \gamma_{ij} \tilde \nabla^l \tilde \nabla_l
\phi\right) - 2 \tilde \gamma_{ij} \tilde \nabla^l \phi \tilde \nabla_l
\phi\right]\,.
\end{eqnarray}

We adopt the \textit{constrained approach} from \cite{Alic:2011a} in
order to enforce the constraints of the conformal formulation; in other
words we enforce the condition ${\rm det} \tilde{\gamma}_{ij}=1$ and
${\rm tr} \tilde{A}_{ij}=0$.

The evolution variable $Z_i$ of the original Z4 formulation is now
included in the ${\hat \Gamma}^i$ variable of the CCZ4 formulation
\begin{equation}
\label{Gammai}
\hat \Gamma^i \equiv \tilde \Gamma^i + 2 \tilde \gamma^{ij} Z_j\,,
\end{equation}
where 
\begin{equation}\label{localGamma}
\tilde \Gamma^i \equiv \tilde \gamma^{jk} \tilde \Gamma^{i}_{jk} =
\tilde \gamma^{ij} \tilde \gamma^{kl} \partial_l \tilde
\gamma_{jk}\,.
\end{equation}
Finally, $\kappa_1$ and $\kappa_2$ are constants associated with the
constraint damping terms and $\kappa_3$ is an extra constant used to
select among different variants of the formulation. In this paper we take
$\kappa_2 = 0$ and $\kappa_3 = 1/2$ and recall that the CCZ4 formulation
is publicly available through the \texttt{Einstein Toolkit}
\cite{Loffler:2011ay, mclachlanweb41}.

The numerical simulations presented in this paper use as gauge conditions
the ``$1+\log$'' slicing \cite{Bona95b}
\begin{equation}
\label{1plog}
  (\partial_t - \beta^i \partial_i) \alpha = -2 \alpha (K - 2 \Theta)\,,
\end{equation}
and the Gamma-driver shift condition \cite{vanMeter:2006vi}
\begin{eqnarray}\label{eq:bssn.gamma.driver}
  (\partial_t - \beta^j \partial_j)\beta^i & = \frac{3}{4} B^i\,, \\
  (\partial_t - \beta^j \partial_j)B^i & = (\partial_t - \beta^j
  \partial_j) \tilde \Gamma^i - \eta B^i\,.
\end{eqnarray}

\subsection{Relativistic Hydrodynamics}

Since we assume the neutron-star matter to be described as a perfect
fluid, the corresponding energy-momentum tensor is given by
\cite{Rezzolla_book:2013}
\begin{equation}
  T_{\mu\nu} = \rho h u_\mu u_\nu + p g_{\mu\nu}\,,
\end{equation}
where $\rho$ is the rest-mass density, $u^\mu$ is the fluid
four-velocity, $p$ is the pressure and $h = 1 + \epsilon + p/\rho$ is the
specific enthalpy. The equations of motion for the fluid are the
``conservation'' of the stress-energy tensor
\begin{equation}\label{eq:euler.equations}
  \nabla_\mu T^{\mu\nu} = 0\,,
\end{equation}
and the baryon number conservation
\begin{equation}
\label{eq:continuity}
  \nabla_\mu (\rho u^\mu) = 0\,.
\end{equation}
These two set of equations are closed by an equation of state (EOS) $p =
p(\rho,\epsilon)$, and we here assume a simple ideal-fluid (or Gamma-law
EOS)
\begin{equation}
p = (\Gamma - 1)\rho\epsilon\,,
\end{equation}
with $\Gamma=2$ case. In some of the tests we also consider the
restriction of this EOS to the isentropic case, \ie the polytropic EOS:
\begin{equation}
p = K \rho^\Gamma\,.
\end{equation}
Although here we use idealized EOSs, the code can make use of the
generic EOS infrastructure developed in the \texttt{Whisky} code and
recently presented in \cite{Galeazzi2013}, which has full support for
thermal and composition-dependent tabulated EOSs.

Equations \eqref{eq:euler.equations} and \eqref{eq:continuity} are solved
in conservation form following the approach by first proposed by
\cite{Banyuls97}, \ie the Valencia formulation, and written as
\begin{equation}\label{eq:valencia}
\frac{\partial\, \pmb{F}^0(\pmb{u})}{\partial t} +
  \frac{\partial\, \pmb{F}^i(\pmb{u})}{\partial x^i} = 
  \pmb{S}(\pmb{u})\,,
\end{equation}
with a vector of primitive variables
\begin{equation}
  \pmb{u} \equiv [\rho, v_i, \epsilon]\,,
\end{equation}
and conservative variables
\begin{equation}
  \pmb{F}^0(\pmb{u}) \equiv \sqrt{\gamma}\, [D,\ S_j,\ \tau]
    = \sqrt{\gamma}\, [\rho W,\ \rho h W^2 v_j,\ \rho h W^2 - p - \rho W]\,.
\end{equation}
The fluxes are 
\begin{equation}
\pmb{F}^i(\pmb{u}) \equiv \sqrt{\gamma} [ D w^i,\ S_j w^i + \alpha p
\delta^i_j,\ \tau w^i + \alpha p v^i ]\,,
\end{equation}
while the sources functions are given by
\begin{equation}
\pmb{S}(\pmb{u}) \equiv \sqrt{\gamma} \bigg[0,\ \frac{\alpha}{2}  S^{lm}
\partial_j \gamma_{lm} + S_k \partial_j \beta^k  - E \partial_j \alpha,
\nonumber \alpha S^{ij} K_{ij} - S^i \partial_i \alpha \bigg]\,.
\end{equation}
In the expressions above the fluid three-velocity measured by the normal
observer is defined as
\begin{equation}
v^i \equiv  \frac{u^i}{\alpha u^0} + \frac{\beta^i}{\alpha}\,,
\end{equation}
while the advection velocity relative to the coordinates is
\begin{equation}
w^i \equiv \alpha v^i - \beta^i\,,
\end{equation}
and the Lorentz factor is defined as $W \equiv \alpha u^0 = (1 - v_i
v^i)^{-1/2}$. Finally, we denoted with $\gamma$ the determinant of the
spatial metric and we have used the fact that $\sqrt{-g} \equiv \alpha
\sqrt{\gamma}$.

\section{THC: A Templated Hydrodynamics Code}
\label{sec:grthc.code}

In this Section we give an overview of \texttt{WhiskyTHC}. First, we
describe the numerical methods used and then a detailed description of
our treatment of fluid--vacuum interfaces, which is one of the key
problems to be addressed in order to attain stable binary evolution,
especially with high-order codes.

\texttt{WhiskyTHC} results from the merger of two codes: \texttt{Whisky}
\cite{Baiotti04} and \texttt{THC} \cite{Radice2012a}. It inherited from
\texttt{THC} the use of high-order flux-vector splitting
finite-differencing techniques and from \texttt{Whisky} the new module
for the recovery of the primitive quantities as well as the new equation
of state framework recently introduced in \cite{Galeazzi2013}. This code
can make use of tabulated, temperature and composition dependent equation
of states, but here we are concerned only with Gamma-law and polytropic
evolutions. More specifically \texttt{WhiskyTHC} solves the equations of
general-relativistic hydrodynamics in conservation form
\eqref{eq:valencia} using a finite-difference scheme. It employs flux
reconstruction in local-characteristic variables using the MP5 scheme,
for which it uses the explicit expression for the eigenvalues and left
and right eigenvectors which can be found in, \eg
\cite{Rezzolla_book:2013}.

The spacetime is evolved using a standard finite-difference method where
all the derivatives, with the exception of the terms associated with the
advection along the shift vector, for which we use a stencil upwinded by
one grid point, are computed with a centered stencil. Typically all these
terms are computed with a fourth-order accurate scheme, but sixth and
eighth order are also available. To ensure the nonlinear stability of the
scheme we add a fifth-order Kreiss-Oliger style artificial dissipation
\cite{Kreiss73}; more details on the code can be found in
\cite{Brown:2008sb}.

Finally, the time evolution and the coupling between the hydrodynamic and
the spacetime solvers is done using the method of lines (MOL), either
with the optimal, strongly-stability preserving third-order Runge-Kutta
scheme, or with the standard fourth-order one.

\subsection{Numerical Methods}
For simplicity we consider, at first, the case of a uniform grid
\begin{equation}
  \bx_{i,j,k} = (i \Delta^1, j \Delta^2, k \Delta^3)\,, \qquad
  i, j, k \in \mathbb{Z}\,.
\end{equation}
The equations of relativistic hydrodynamics \eqref{eq:valencia} are
written on such a grid using the method of lines in a semidiscrete,
dimensionally unsplit way as
\begin{eqnarray}
\label{eq:mol}
\fl \frac{d \bF^0_{i,j,k}}{d t} = 
  \bS_{i,j,k}  +
  \frac{\bF^1_{i-1/2,j,k} - \bF^1_{i+1/2,j,k}}{\Delta^1}
   + \frac{\bF^2_{i,j-1/2,k} - \bF^2_{i,j+1/2,k}}{\Delta^2} + \nonumber \\
 \phantom{+}  \frac{\bF^3_{i,j,k-1/2} - \bF^3_{i,j,k+1/2}}{\Delta^3}\,,
\end{eqnarray}
where $\psi_{i,j,k}$ is the value of a generic quantity, $\psi$, at
$\bx_{i,j,k}$, while $(\bF^1_{i-1/2,j,k} - \bF^1_{i+1/2,j,k} ) /
\Delta^1$ is a high-order, non-oscillatory, approximation of $-\partial
\bF^1 / \partial x^1$ at $\bx_{i,j,k}$, whose explicit expression still
needs to be specified.

To illustrate how we compute the discrete derivatives on the
right-hand-side of (\ref{eq:mol}) it is useful to make a step back and
consider, first a simpler scalar hyperbolic equation in one dimension,
\ie
\begin{equation}\label{eq:scalar.claw}
  \frac{\partial u}{\partial t} + \frac{\partial f(u)}{\partial x} = 0\,.
\end{equation}
We introduce a uniform grid, $x_i = i \Delta$, and define, for any function,
$v(x)$ the volume averages
\begin{equation}
\label{eq:average}
  \tilde{v}_i \equiv \frac{1}{\Delta} \int_{x_{i-1/2}}^{x_{i+1/2}} v(x)\,
  d x\,.
\end{equation}

A reconstruction operator, $\mathcal{R}$, is a nonlinear operator
yielding a high-order approximation of $v$ at a given point $x$ using
its volume averages, $\tilde{v}_i$. Since $v(x)$ can be discontinuous, we
distinguish two different reconstruction operators, $\mathcal{R}^-$ and
$\mathcal{R}^+$, called the left-biased and right-biased reconstruction
operators, such that
\begin{eqnarray}
 & \left[\mathcal{R}^-\left(\{\tilde{v}_i\}\right)\right](x) =
    \lim_{y \to x^{-}} v(y) + \mathcal{O}(\Delta^r)\,, \\
 & \left[\mathcal{R}^+\left(\{\tilde{v}_i\}\right)\right](x) =
    \lim_{y \to x^{+}} v(y) + \mathcal{O}(\Delta^r)\,,
\end{eqnarray}
where we have used the notation $\mathcal{R}^+ \left(\{\tilde{v}_i
\}\right)$ to remark that $\mathcal{R}$ is an operator that acts on a set
of averages $\tilde{v}_i$, and where $r$ is the order of the
reconstruction operator $\mathcal{R}$. Hereafter we will use the notation
$v_{i+1/2}^-$ and $v_{i+1/2}^+$ to denote the reconstructed values in
$x_{i+1/2}$ using $\mathcal{R}^-$ and $\mathcal{R}^+$ respectively, \ie.
\begin{eqnarray}
 & v_{i+1/2}^- \equiv \left[
\mathcal{R}^-\left(\{\tilde{v}_{i}\}\right)\right](x_{i+1/2})\,,\\
 & v_{i+1/2}^+ \equiv \left[
\mathcal{R}^+\left(\{\tilde{v}_{i}\}\right)\right](x_{i+1/2})\,.
\end{eqnarray}

Our code implements a wide range of reconstruction operators, from the
second-order minmod to the seventh-order weighted essentially
non-oscillatory (WENO) reconstructions, but all the calculations in this
paper were performed using the fifth-order monotonicity preserving (MP5)
scheme, which, as discussed in \cite{Radice2012a}, represent a good
compromise between robustness and accuracy.

The reconstruction operators are the core components of both
finite-volume and finite-difference schemes. In a finite-volume scheme
they are used to compute the left and right state to be used in the
(usually approximate) Riemann solver to compute the fluxes (see, \eg
\cite{Rezzolla_book:2013} for details). In a finite-difference scheme,
instead, they are used to compute the above mentioned non-oscillatory
approximation of $\partial f / \partial x$. Following \cite{Shu88} we
introduce a function $h(x)$ and such that
\begin{equation}
\label{eq:h_of_x_1}
  f\big[u(x_i)\big] = \frac{1}{\Delta} \int_{x_{i-1/2}}^{x_{i+1/2}}
  h(\xi)\, d\xi\,,
\end{equation}
that is, the average of $h(x)$ between $x_{i-1/2}$ and $x_{i+1/2}$
corresponds to the value of $f$ at $x_{i}$. Next, we note that
\begin{equation}
\label{eq:h_of_x_2}
  \frac{\partial f}{\partial x}\bigg|_{x_i} =
  \frac{h(x_{i+1/2})-h(x_{i-1/2})}{\Delta}=
  \frac{h_{i+1/2}-h_{i-1/2}}{\Delta}\,,
\end{equation}
where both~\eqref{eq:h_of_x_1} and \eqref{eq:h_of_x_2} are exact
expressions. Hence, by using the usual reconstruction operators
$\mathcal{R}$ of order $r$ to reconstruct $h_{i+1/2}$, one obtains a
corresponding accurate approximation of order $r$ of the derivative
$\partial f / \partial x$ at $x_{i}$. Note that $h$ is never actually
computed at any time during the calculation as we only need the values
of $f$ at the grid points, \ie $f\big[u(x_i)\big]$.

In order to ensure the stability of the resulting scheme, one has to take
care to upwind the reconstruction appropriately. Let us first consider
the case in which $f'(u) \equiv \partial f /\partial u > 0$. If we set
\begin{equation}
  \tilde{v}_i = f\big[u(x_i)\big] = \frac{1}{\Delta}
  \int_{x_{i-1/2}}^{x_{i+1/2}} h(\xi)\, d\xi\,,
\end{equation}
and 
\begin{equation}
f_{i+1/2} \equiv v_{i+1/2}^-\,,\qquad f_{i-1/2} \equiv v_{i-1/2}^-\,,
\end{equation}
then 
\begin{equation}
  \frac{\partial f(u)}{\partial x} = \frac{f_{i+1/2} - f_{i-1/2}}{\Delta}
    + \mathcal{O}(\Delta^r)\,,
\end{equation}
gives the wanted high-order approximation of $\partial f / \partial x$
at $x_i$.

In the more general case, where the sign of $f'(u)$ needs to be
determined, in order to compute $f_{i+1/2}$, we have to split $f$ in a
right-going flux, $f^+$, and a left-going one, $f^-$, \ie $f = f^+ +
f^-$, and use the appropriate upwind-biased reconstruction operators
separately on both parts, in order to guarantee the stability of the
method.

There are several different ways to perform such a split and in our
code we implemented two of them: the Roe flux-split, \ie
\begin{equation}\label{eq:roe.flux.split}
f = f^{\pm}, \quad \textrm{if } [f'(\bar{u})]_{x_{i+1/2}} \gtrless 0\,,
\end{equation}
where $\bar{u}_{i+1/2} \equiv \frac{1}{2} (u_i + u_{i+1})$, and the
Lax-Friedrichs or Rusanov flux-split \cite{Shu97}, \ie
\begin{equation}\label{eq:lax.friedrichs.split}
  f^\pm = f(u) \pm \alpha u, \quad \alpha = \max [f'(u)]\,,
\end{equation}
where the maximum is taken over the stencil of the reconstruction
operator. The Roe flux-split is less dissipative and yields a
computationally less-expensive scheme, since only one reconstruction is
required instead of two, but its use can result in the creation of
entropy-violating shocks in the presence of transonic rarefaction waves
(see, \eg \cite{LeVeque92}), and it is also susceptible to the carbuncle
(or odd-even decoupling) phenomenon \cite{Quirk1994}. To avoid these
drawbacks, we switch from the Roe to the Lax-Friedrichs flux split when
$u$ or $f$ are not monotonic within the reconstruction stencil.

We now go back to the more general system of equations
(\ref{eq:valencia}). The derivatives $\partial \bF^a_{i,j,k} / \partial
x^a$ can be computed following the procedure outlined above on a
component-by-component basis. This approach is commonly adopted in the
case of low-order schemes, but it often results in spurious numerical
oscillations in the high-order (usually higher then second) case. To
avoid this issue the reconstruction should be performed on the local
characteristic variables of the systems. To avoid an excessively
complex notation, let us concentrate on the fluxes in the
$x$-direction; in this case, to reconstruct $\bF^1_{i+1/2,j,k}$, we
introduce the Jacobian matrices
\begin{equation}
\label{eq:jacobians}
  \boldsymbol{A}^\alpha = \frac{\partial \bF^\alpha}{\partial \bu}
    \bigg|_{\bar{\bu}}\,, \qquad \qquad \alpha = 0, 1,
\end{equation}
where $\bar{\bu}$ is an average state at the point where the
reconstruction is to be performed. In our code $\bar{\rho}$ and
$\bar{\epsilon}$ are computed as a simple average of the left and right
states \cite{Radice2012a}, while, in order to avoid creating unphysical
values in the velocity, $\bar{v}^i$ is computed from the averages of $W
v^i$ in the left and right states.

Hyperbolicity of (\ref{eq:valencia}) implies that $\boldsymbol{A}^0$ is
invertible and the generalised eigenvalue problem
\begin{equation}
[\boldsymbol{A}^1 - \lambda_{(I)} \boldsymbol{A}^0] \boldsymbol{r}_{(I)} = 0\,,
\end{equation}
has only real eigenvalues, $\lambda_{(I)}$, and $N$ independent, real
right-eigenvectors, $\boldsymbol{r}_{(I)}$ \cite{Anile_book} (see, \eg
\cite{Font08 ,Rezzolla_book:2013} for the explicit expressions of the
eigenvalues and eigenvectors). We will denote with $\boldsymbol{R}$ the
matrix of right eigenvectors, \ie
\begin{equation}
  R^I_{J}  = r^I_{(J)}\,,
\end{equation}
and with $\boldsymbol{L}$ its inverse. We define the local
characteristic variables
\begin{equation}
  \boldsymbol{w} = \boldsymbol{L}\, \bu\,, \qquad \boldsymbol{Q} =
  \boldsymbol{L}\,\bF^1 \,,
\end{equation}
and compute $\boldsymbol{Q}_{i+1/2,j,k}$ doing a component-wise
reconstruction, where $\boldsymbol{w}$ is used in place of $u$ and
$\boldsymbol{Q}$ in place of $f$ in Eq.
(\ref{eq:lax.friedrichs.split}). Finally, we set
\begin{equation}
  \bF^1_{i+1/2,j,k}  = \boldsymbol{R}\, \boldsymbol{Q}_{i+1/2,j,k}\,.
\end{equation}
This procedure is repeated in the other directions and yields the wanted
approximations of the $\partial \bF^a / \partial x^a$ terms in
$\bx_{i,j,k}$. In all of the results presented here the reconstruction is
typically performed in local characteristic variables. Exceptions are
low-density points (at least for some of the atmosphere prescriptions;
more on this below) or points which, on the basis of the value of the
lapse function, we estimate being inside an apparent horizon. In these
cases we switch to the component-wise reconstruction of the fluxes with
Lax-Friedrichs flux-split.

The primitive variables are recovered after each sub-step using a
numerical root-finding procedure. In particular \texttt{WhiskyTHC} uses
the EOS and primitive variables recovery framework recently developed for
the \texttt{Whisky} code and described in details in \cite{Galeazzi2013}.
The only differences between our approach and the one discussed there is
that 1) we do not alter the conservative quantities to match the
primitive ones in the case in which adjustments have been made, 2) to
ensure the well behaviour of the eigenvectors (which become degenerate if
the sound speed goes to zero) we enforce a floor on $\tau$ (more
on this in the next Section). The reason why we do not alter the
conserved quantities in case of failures is that we prefer to rely on the
evolution to bring them back into the physically allowed region, instead
of changing them in an arbitrary manner. We also note that any adjustment
made in the recovery can be simply interpreted as an error in the
calculation of the flux/source terms

Finally, our code has been tested using a rather basic subset of the
features supported by \texttt{Carpet} \cite{Schnetter-etal-03b}, the AMR
driver of the \texttt{Cactus} computational toolkit \cite{Goodale02a} on
top of which our code is built. \texttt{Carpet} supports
Berger-Oliger-style mesh refinement \cite{Berger84, Berger89,
Reisswig2012b} with sub-cycling in time and refluxing. At the moment, our
code has been tested only with static grid refinement. It supports
sub-cycling in time, but not refluxing. While we have plans to implement
refluxing and test high-order prolongation operators in order to use
dynamical grids, we also note that these features are not of fundamental
importance for the study of gravitational waves from inspiraling binary
neutron stars, which is the main aim of our code. 

\subsection{Treatment of Vacuum Regions}

The treatment of interfaces between vacuum region and fluid regions is
one of the most challenging problems in Eulerian (relativistic)
hydrodynamics codes (see \eg~\cite{galeazzi_master, kastaun_2006_hrs,
  Millmore2010}), especially when studying near-equilibrium
configuration, such as an oscillating compact star, having large density
gradients close to the surface and over long timescales. The most
commonly used approach to treat vacuum regions is to fill them with a
low-density fluid, the ``atmosphere'', such that if a fluid element is
evolved to have a rest-mass density below a certain threshold, it is set
to have a floor value and zero coordinate
velocity~\cite{Font02c,Baiotti04}. This approach works reasonably well
for standard second-order codes and has been adopted by the vast majority
of the relativistic-hydrodynamics codes, but it is problematic for
higher-order codes \cite{Radice2011}. The reason is that small numerical
oscillations can easily result in the creation of very low-density
regions that with the prescription for the floor, violating the
conservative character of the equations and creating artificially mass,
energy and momentum. As a result, they are subsequently amplified,
ultimately destabilizing the evolution. The situation is even more
complicated for a code, such as ours, which relies on characteristic
variables as they become degenerate in the low-density, low-temperature
limits.

We notice that for many applications, such as the inspiral of binary
neutron stars (at least up to contact), or the oscillation of single
stars, the treatment of the stellar surface is one of the main challenges
and the only reason why low-order, but robust shock capturing codes are
commonly used. Indeed the problem of the vacuum treatment is one of the
main obstacles on the road to high-order general-relativistic
hydrodynamics codes. For this reason, it is instructive to address this
problem in detail as we do in the following.

However, before going into the details of the treatment, it is useful to
make two rather general remarks. First, we should point out that the MP5
scheme is remarkably robust even in conjunction with the most basic
atmosphere treatment that we implemented, \ie one in which no additional
modification is made on the scheme at the interface between vacuum and
fluid region, beside the imposition of a minimum rest-mass density level (more on
this below). In our preliminary tests, other schemes, such as WENO5,
which do not enforce the monotonicity of the reconstruction, could not
yield stable evolutions even for single stars in the Cowling
approximation. Second, most of the problems with the atmosphere appear in
points where the surface of the star is aligned with the grid, because
along these directions the numerical dissipation is minimal. These
artefacts, that we discuss in more detail in the next Section, are easily
``fixed'' with the use of extra numerical dissipation close to the
surface of the star. Keeping this in mind, we now give the details of the
three different prescriptions that we developed for the treatment of the
low-density regions.

\subsubsection{Standard Atmosphere Treatment}

The first prescription is what we call the ``ordinary MP5'' approach. It
follows the lines of what is most commonly done to treat vacuum in
general-relativistic hydrodynamics. First, we choose a minimum rest-mass
density $\rho_{\mathrm{atmo}}$, which we take to be, typically, in the
range $(10^{-7} - 10^{-9}) \rho_{\mathrm{ref}}$, $\rho_{\mathrm{ref}}$
being some reference rest-mass density (normally the initial maximum
rest-mass density). Second, we choose a tolerance parameter,
$\varepsilon$, typically $10^{-2}$, chosen to avoid excessive
oscillations of the fluid--vacuum interface so that points where the
rest-mass density falls below $(1+\varepsilon)\rho_{\mathrm{atmo}}$, are
set to atmosphere. In particular, the rest-mass density is set to
$\rho_{\mathrm{atmo}}$, the velocity to zero and the internal energy is
calculated assuming a polytropic EOS. In addition, we enforce a floor for
the conserved energy density $\tau$, $\tau_{\mathrm{atmo}} =
\rho_{\mathrm{atmo}} \epsilon_{\mathrm{atmo}}$.

As we show below, this approach, is already perfectly adequate for
inspiraling binary neutron stars, but it might have problems in the case
of slowly moving vacuum-fluid interfaces aligned with the grid,
especially in the case of isentropic evolutions where the surface remains
a sharp interface and no spurious heating can occur.

\subsubsection{Improved Atmosphere Treatment}

In order to improve our atmosphere treatment, we introduced an
alternative method in which we increase the level of dissipation of the
scheme by switching to the component-wise Lax-Friedrichs flux split below
a certain rest-mass density. Typical values for this new threshold are
chosen so that the first one or two grid points in the stellar interior
are evolved using the Lax-Friedrichs flux split. The use of
component-wise reconstruction, as opposed to characteristic-wise, is done
to avoid problems due to the degeneracy of the characteristic variables
close to vacuum and to avoid polluting quantities, such as the linear
momentum, with the numerical errors present in the internal energy (which
is typically large in the immediate vicinity of the atmosphere). This is
what we refer to as ``MP5+LF'' approach.

This latter approach is more robust, but can also result in the creation
of artefacts in the case in which low-density matter is ejected from the
stellar surface. In this case, the fluid typically presents a rather
smooth interface with vacuum, so that one would expect to be able to
treat it with high accuracy. Unfortunately, as we show in the next
Section, if the rest-mass density of the ejecta falls below the
Lax-Friedrichs threshold, the use of a component-wise reconstruction
yields a rather oscillatory solution, with the creation of patchy regions
of lower/higher rest-mass density in very dynamical situations (\cf
second panel in Figure~\ref{fig:grthc.migration.rho2d} and the discussion
there).

\subsubsection{Positivity Preserving Limiter}

Overall neither of the previous methods is completely satisfactory.  For
this reason we propose here a novel approach based on the use of the
positivity preserving limiter recently proposed by \cite{Hu2013}, which
is significantly simpler to implement than the ``classical''
positivity-preserving limiters already proposed in the literature, \eg
\cite{Zhang2010, Zhang2011, Zhang2011a, Balsara2012}.

For the sake of completeness, we give here a brief overview of the key
ideas presented in \cite{Hu2013}, to which we refer to for a more
complete presentation. To keep the notation simple we consider, at first,
a scalar conservation law in one dimension
\begin{equation}\label{eq:grthc.pp.example}
  \frac{\partial u}{\partial t} + \frac{\partial f(u)}{\partial x} = 0\,.
\end{equation}
We notice that any scheme able to guarantee the positivity of $u$ over
one first-order Euler timestep, will automatically guarantee positivity
when used with any SSP time integrator, as in these schemes the time
update is always constructed as a convex combination\footnote{We recall
  that a convex combination of a set of vectors,
  $\boldsymbol{\vec{x}}_i$, is a combination of the form $\sum_i c_i
  \boldsymbol{\vec{x}}_i$, where $0 \leq c_i \leq 1$ and $\sum_i c_i =
  1$.} of Euler steps. For this reason we consider a discretization of
\eqref{eq:grthc.pp.example} of the form
\begin{equation}
\label{eq:grthc.pp.method}
  \frac{u^{n+1}_i - u^n_i}{\Delta^0} = \frac{f_{i-1/2} -
  f_{i+1/2}}{\Delta^1}\,.
\end{equation}
If we let $\lambda \equiv \Delta^0/\Delta^1$, then the previous can be
written as
\begin{equation}
\label{eq:grthc.pp.split}
u^{n+1}_i = \frac{1}{2} (u^+_i + u^-_i)
          = \frac{1}{2} \big[ (u_i^n + 2 \lambda f_{i-1/2} ) + (u_i^n - 2
          \lambda f_{i+1/2}) \big]\,.
\end{equation}
where $u^+_i = u_i^n + 2 \lambda f_{i-1/2}$ and $u^-_i = u_i^n - 2
\lambda f_{i+1/2}$. Clearly, if $u^+_i$ and $u^-_i$ are positive, so will
be $u^{n+1}_i$. The key observation made in \cite{Hu2013} is that, if
$f_{i+1/2}$ and $f_{i-1/2}$ are computed with the first-order
Lax-Friedrichs scheme with $\lambda \leq {1}/{2a}$, $a$ being the
largest propagation speed, then $u^+_i, u^-_i \geq \min_i (u^n_i)$
\cite{Zhang2010}.

The idea is to modify Eq.~\eqref{eq:grthc.pp.method} as
\begin{equation}\label{eq:grthc.pp.flux}
  f_{i+1/2} = \theta f_{i+1/2}^{\mathrm{HO}} + (1-\theta)
  f_{i+1/2}^{\mathrm{LF}}\,,
\end{equation}
where $f_{i+1/2}^{\mathrm{HO}}$ is the high-order flux of the original
scheme, $f_{i+1/2}^{\mathrm{LF}}$ is the flux associated with the
first-order Lax-Friedrichs scheme, and $\theta \in [0,1]$ is chosen to be
the maximum value such that both $u^-_i$ and $u^+_{i+1}$ are positive. In
regions where the solution is far from vacuum, $\theta=1$, so that the
high-order fluxes are used (and in particular the formal order of
accuracy of the scheme remains unchanged). In regions close to vacuum, it
is always possible to find some $\theta \geq 0$ such that positivity is
guaranteed, since for $\theta = 0$ the scheme reduces to the
Lax-Friedrichs scheme, which is known to be positivity preserving.

The multi-dimensional extension of Eq.~\eqref{eq:grthc.pp.split} is done
in a component-by-component fashion. For instance in three dimensions
\eqref{eq:grthc.pp.split} becomes

\begin{eqnarray}
\label{eq:grthc.pp.split3}
 \fl 
u^{n+1}_{i,j,k} =
  && \hskip -1.3cm
\frac{\alpha_x}{2} \Bigg[ \bigg(u^n_{i,j,k} +
      2 \frac{\lambda_x}{\alpha_x} f_{i-1/2,j,k}\bigg) +
     \bigg(u^n_{i,j,k} - 2 \frac{\lambda_x}{\alpha_x}
     f_{i+1/2,j,k}\bigg)\Bigg] + \nonumber \\
  && \hskip -1.3cm 
\frac{\alpha_y}{2} \Bigg[ \bigg(u^n_{i,j,k} +
      2 \frac{\lambda_y}{\alpha_y} f_{i,j-1/2,k}\bigg) +
    \bigg(u^n_{i,j,k} - 2 \frac{\lambda_y}{\alpha_y}
    f_{i,j+1/2,k}\bigg)\Bigg] + \nonumber \\
  && \hskip -1.3cm
\frac{\alpha_z}{2} \Bigg[ \bigg(u^n_{i,j,k} +
      2 \frac{\lambda_z}{\alpha_z} f_{i,j,k-1/2}\bigg) +
    \bigg(u^n_{i,j,k} - 2 \frac{\lambda_z}{\alpha_z}
    f_{i,j,k+1/2}\bigg)\Bigg]\,,
\end{eqnarray}
where $\alpha_x, \alpha_y, \alpha_z$ are positive constants such that
$\alpha_x + \alpha_y + \alpha_z = 1$, typically chosen so as to be equal
to $1/3$ (but see the remarks at the end of the Section). The limiter at
each interface is then chosen enforcing positivity of the terms in the
round brackets.

This approach can then be easily extended to systems of equations
\cite{Hu2013}. In particular, \cite{Hu2013} constructed a limiter
able to guarantee the positivity of rest-mass density and pressure for
the classical equations of hydrodynamics also in the case in which source
terms are present\footnote{Note that in this case a smaller timestep
  might be required, depending on the nature of the source terms}.

In the general-relativistic case, it is not trivial to enforce the
positivity of the pressure, especially for tabulated EOS, because of the
presence of complex source terms in the energy equations. For this
reason, as was the case for the atmosphere treatment, we need to enforce
positivity of the pressure with the imposition of a floor on $\tau$. On
the other hand, the continuity equation
\begin{equation}
  \partial_t \hat D + \partial_j [\hat D w^j] = 0\,,
\end{equation}
where $\hat D \equiv \sqrt{\gamma} \rho W$ and $w^j \equiv \alpha v^j -
\beta^j$, is formally equivalent to the Newtonian continuity equation
after the identification
\begin{eqnarray}
  \hat D \longleftrightarrow \rho\,, \qquad \qquad
  w^i \longleftrightarrow v^i\,,
\end{eqnarray}
thus one can construct a scheme ensuring the positivity of $\hat D$ by
simply adopting the prescriptions used by \cite{Hu2013} to guarantee the
positivity of the density for the Newtonian Euler equations.

Some comments should be made on the positivity preserving limiter. First,
the positivity preserving limiter is not directly a way to treat
vacuum--fluid interfaces in a physically accurate way, for the simple
reason that the fluid model is not adequate to represent such
transitions. A proper modeling of the stellar surface can only be done by
treating it as a free boundary of the problem determined by the balance
between inertial and gravitational forces on the fluid as done, for
instance, in \cite{kastaun_2006_hrs}. For this reason its use does not
free us from having a low-density fluid everywhere or from having to
manually enforce that $\hat D > \hat D_{\mathrm{atmo}}$. This may be
necessary because in some situations, for instance at the surface of a
star, the high-order fluxes and the Lax-Friedrichs fluxes can differ by
several orders of magnitude, so that small floating-point errors can
drift the conserved density, $\hat D$, below the minimum. This is done by
simply resetting $\hat D$ to $\hat D_{\mathrm{atmo}}$ whether $\hat D <
\hat D_{\mathrm{atmo}}$, without changing the other quantities.

What the positivity preserving limiter does, however, is to ensure local
conservation of the solution up to floating-point precision. This is
because the floor for the fluid's rest-mass density can be
\emph{arbitrarily small} and does not require any tuning. As a result,
de-facto it prevents the scheme from extracting/losing mass from/to the
atmosphere because of numerical oscillations. In contrast, the classical
atmosphere prescriptions usually work only in a limited range of
$\rho_{\mathrm{atmo}}$ and $\varepsilon$ as these coefficients must be
tuned in order to achieve a balance between the amount of mass extracted
from the atmosphere (which typically increases as $\rho_{\mathrm{atmo}}$
decreases) and the mass lost into it (which typically increases as
$\varepsilon$ increases). With the positivity preserving limiter instead,
in situations where we need to reset $\hat D$, we are guaranteed that
this correction is of the order of the floating-point precision with
respect to the typical rest-mass densities we are actually interested in
tracking.

The way in which we use the positivity preserving limiter is
rather simple: we fill the vacuum with a low rest-mass density floor at
the beginning of a simulations and we let it evolve freely, only relying
on the positivity preserving limiters to ensure its proper
behaviour. This typically results in the creation of accretion flows onto
our compact objects. However, given the low rest-mass density of the
floor, which we take to be $\sim 10^{-16} \rho_{\mathrm{ref}}$ (\ie below
floating-point precision!), the effects of this artificial accretion are
completely negligible. 

To avoid problems with the decomposition of the Jacobian in eigenvalues
and eigenvectors, we also switch to component-wise reconstruction below a
certain rest-mass density, typically $10^{-7} \rho_{\mathrm{ref}}$, but
this, in contrast to the prescription outlined in the previous Section,
has little dynamical effect as flows at those densities are, anyway,
completely dominated by numerical effects.  Moreover, as we show in the
next Section, even if the floor rest-mass density is taken to be
unnecessarily large, the use of positivity preserving limiters results in
much smaller perturbations with respect to the use of a more traditional
atmosphere treatment.

Finally, a comment concerning the timestep constraint, as this may be seen
as the only real limitation of the positivity preserving approach. For
the scheme to ensure positivity in the multi-dimensional case, one must
ensure $a \alpha_{i} {\Delta^0}/{\Delta^i} < 1$. Since $a \sim 1$ and
$\alpha_{i} = 1/D$, $D$ being the number of dimensions, this results in a
rather stringent Courant-Friedrichs-Lewy (CFL) condition. In practice we
find our scheme to be robust even for much larger timesteps, probably
because the advection velocity in the low rest-mass density regions is
typically smaller than the maximum velocity and because Lax-Friedrichs
scheme is actually positivity preserving even with Courant factor
$\mathrm{CFL}=1$ in one dimension \cite{Zhang2011a} [$\mathrm{CFL}=1/D$
  in $D$ dimensions, even though it is not possible to guarantee that
  $u_i^+$ and $u_{i}^-$ in Eq. \eqref{eq:grthc.pp.split} are separately
  positive]. In order to use larger timesteps, we simply compute the
value of the limiter $\theta$ assuming $\alpha_{i} = 1$, as in the
one-dimensional case (note this \textit{does not mean} that we evolve
using \eqref{eq:grthc.pp.split3} with the $\alpha_{i}$'s equal to one),
and we set it to zero (\ie we use Lax-Friedrichs fluxes) when it is not
possible to enforce the positivity of $u_i^+$ and $u_{i+1}^-$. We have
found this procedure to be sufficient to prevent negative densities from
occurring (at least at to a reasonable extent) and to be computationally
much less expensive with respect to the approach in which $\alpha_i =
1/D$.

\section{Isolated Neutron Stars}
\label{sec:grthc.sns}

In this Section we describe a series of representative tests that we
performed with \texttt{WhiskyTHC} in the case of single, isolated,
nonrotating neutron stars (or TOVs from Tolman-Oppenheimer-Volkoff
\cite{Tolman39,Oppenheimer39b}). First, we present the results obtained
in the Cowling approximation, \ie without evolving the spacetime, for
perturbed, oscillating stars. Then we proceed to analyze the case of
linear oscillations of stable stars in full general relativity
(GR). Finally, we show the results obtained for the evolution of unstable
stars: both for the migration and for the collapse to a black hole. The
focus of our discussion is mostly on the effects of the different
prescriptions for the treatment of the atmosphere. Hereafter we will
denote the basic treatment as ``MP5'', the enhanced treatment with extra
dissipation on the surface as ``MP5+LF'', and the positivity preserving
treatments as ``MP5+PP''.

\subsection{Linear Oscillations: Cowling Approximation}
\label{sec:grthc.cowling}

The first test that we consider is the long-term evolution of a
perturbed, isolated, nonrotating, neutron star in the Cowling
approximation. The goal of this test is to assess the impact on the
accuracy of the three different atmosphere treatments over long
timescales.  We consider a model described by the polytropic EOS with
$K=100$ and $\Gamma=2$. The initial central rest-mass density is
$\rho_c(0) = 1.28\times 10^{-3}\ M_\odot^{-2}$, yielding a model with an
ADM mass of $1.4\ M_\odot$. The initial velocity is perturbed with the
injection of a radial eigenfunction that is exact in the Cowling
approximation. The maximum amplitude of the perturbation is $|v^r| \simeq
0.024$ and the initial perturbation is ingoing.

We evolve this model for $10,000\ M_\odot$, \ie $\simeq 130$ dynamical
timescales, using our different prescriptions for the atmosphere. Our
fiducial resolution is $h = 0.2\ M_\odot$ so that the radius of the star
is covered with $\simeq 45$ grid points. In order to make a fair
comparison, we use the same atmosphere threshold for all the methods,
$\rho_{\mathrm{atmo}} = 10^{-10}\ M_\odot^{-2}$, and evolve all the
models with the third-order SSP-RK3 with $\mathrm{CFL} = 0.4$. The
gravitational source terms are computed using sixth-order finite
differencing. Finally, the evolution is computed only in the octant $x, y,
z \geq 0$ and we assumed reflection symmetry across the $(x,y),\, (x,z)$
and $(y,z)$ planes.

\begin{figure}
\begin{center}
  \includegraphics[width=0.7\textwidth]{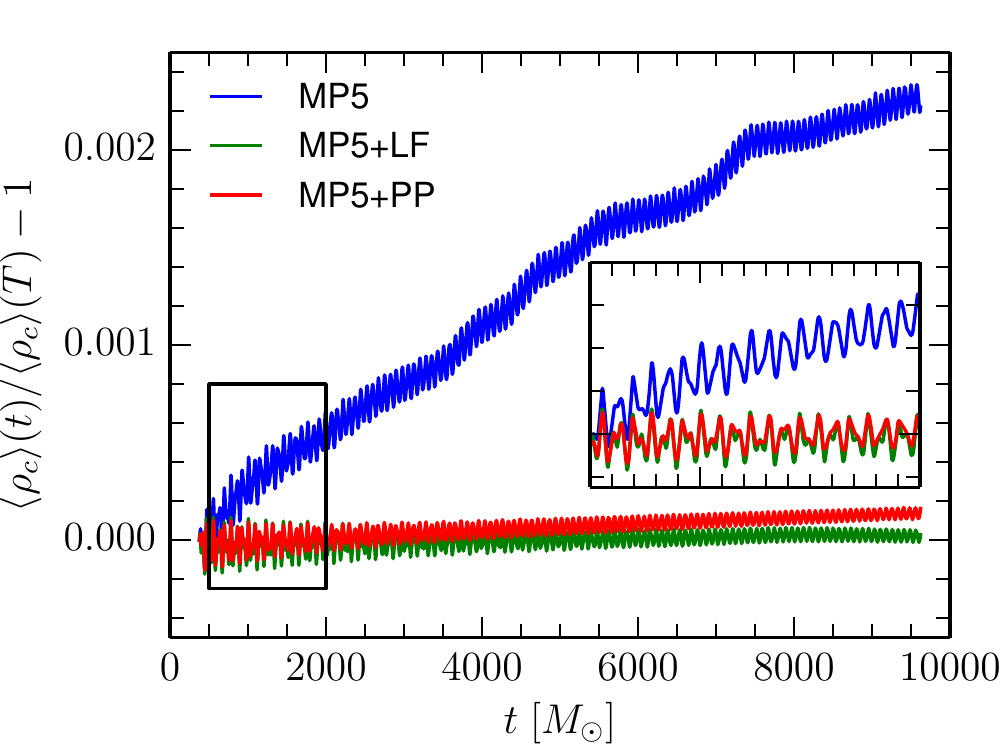}
  \caption{\label{fig:grthc.cowling.rho} Normalized central rest-mass
  density for the perturbed TOV in the Cowling approximation and for
  different atmosphere prescriptions. The inset shows a magnification of
  the dynamics in the time interval $[500, 2000]\, M_\odot$, \ie the area
  bounded by the black rectangle.}
\end{center}
\end{figure}

The evolution of the central rest-mass density, $\rho_c$, is shown in
Figure  \ref{fig:grthc.cowling.rho}. In particular, in order to highlight
the secular trend of the data, we show a moving average of $\rho_c$
defined as
\begin{equation}
  \langle \rho_c \rangle (t) \equiv \frac{1}{2T} \int_{t-T}^{t+T} \rho_c(t')\,
  \dd t', \qquad T \leq t \leq T_{\mathrm{final}} - T,
\end{equation}
where $T \equiv 5 / f_F$, $f_F$ being the frequency of the $F$-mode from
linear perturbation theory. All the different prescriptions yield very
similar evolutions of the central rest-mass density which presents a
series of slowly damped oscillations. The pulsation frequency agrees, to
within the nominal error of the discrete Fourier transform, with the one
expected from linear perturbation theory in the Cowling
approximation. The power-spectrum also shows small contributions from
higher-order overtones (\ie more than a factor ten smaller than the
$F$-mode) as well as an even smaller nonlinear component at integer
multiples of the $F$-mode frequency. In the case of the MP5+LF
prescription, we verified that the nonlinear component decreases with
decreasing perturbation amplitudes and that it is not distinguishable
from the background noise for perturbation amplitudes $\simeq 2\times
10^{-3}$. The ordinary MP5 prescription also shows a small secular
increase in the central rest-mass density. Apart from this, all the
schemes appear to be able to yield very clean oscillations.

\begin{figure}
\begin{center}
  \includegraphics[width=0.7\textwidth]{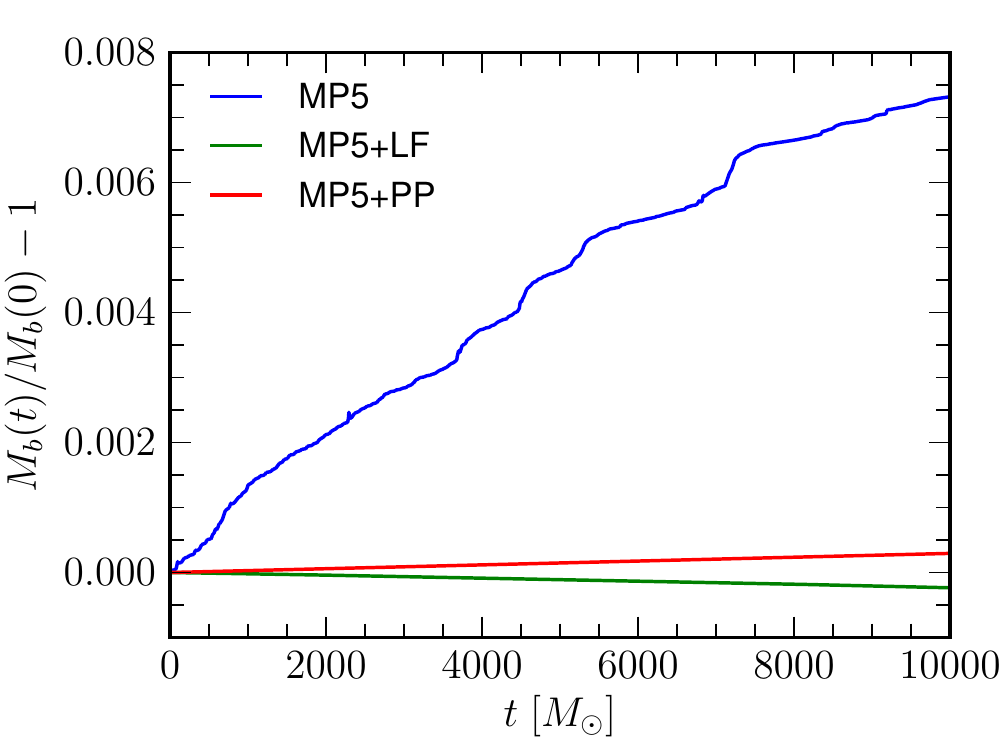}
  \caption{\label{fig:grthc.cowling.restmass}
  Normalized total rest-mass variations for the perturbed TOV in the
  Cowling approximation and for different atmosphere prescriptions.
  }
\end{center}
\end{figure}

The difference between the different schemes can be better appreciated by
looking at Figure \ref{fig:grthc.cowling.restmass} where we show the
evolution of the total rest mass
\begin{equation}
  M_b(t) = \int_{\Sigma_t} \rho\, \sqrt{\gamma}\, \dd^3 x\,,
\end{equation}
for the different models. Overall, the rest-mass conservation is at
acceptable levels for all the methods, \ie $\Delta M_b/M_b \lesssim
10^{-3}$, but the ordinary MP5 prescription is clearly the one with the
largest error, as it shows larger variations with respect to the other
schemes. This, in turn, is responsible for the secular drift mentioned
earlier (an increase in the total rest mass leads to an increase in the
central rest-mass density). The Lax-Friedrichs flux-switch at the
surface, instead, results in a steady loss of matter which is slowly
diffused into the atmosphere, while the MP5+PP approach yields a steady
increase in the rest mass because of the accretion of the low-density
floor which is continuously ``injected'' from the outer boundary (we
simply fix the rest-mass density in the ghost regions at the outer
boundary to its initial value).

\begin{figure*}
  \begin{minipage}{0.5\hsize}
    \begin{center}
      \includegraphics[width=1.0\textwidth]{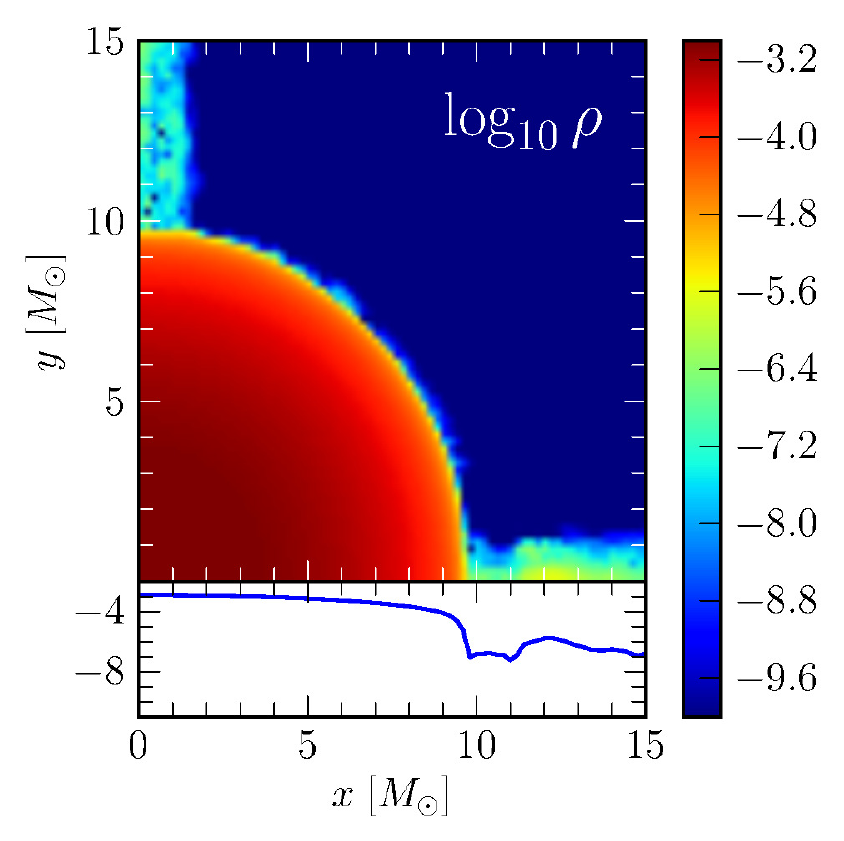}
      \\ MP5
    \end{center}
  \end{minipage}
  \begin{minipage}{0.5\hsize}
    \begin{center}
      \includegraphics[width=1.0\textwidth]{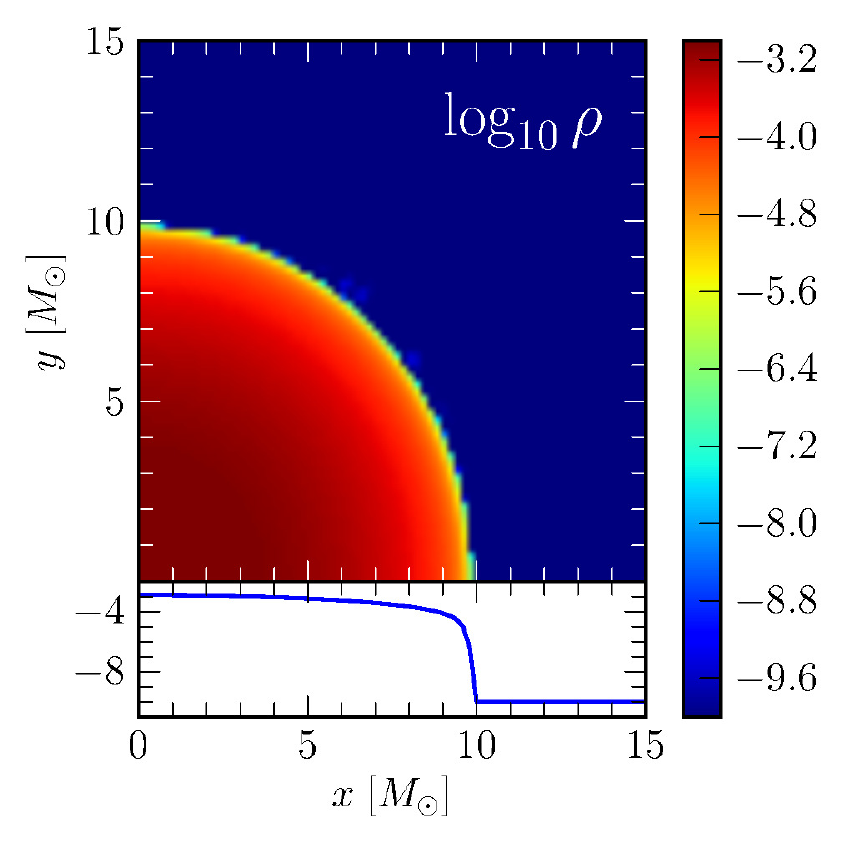}
      \\ MP5+LF
    \end{center}
  \end{minipage}
  \begin{center}
  \begin{minipage}{0.5\hsize}
      \begin{center}
      \includegraphics[width=1.0\textwidth]{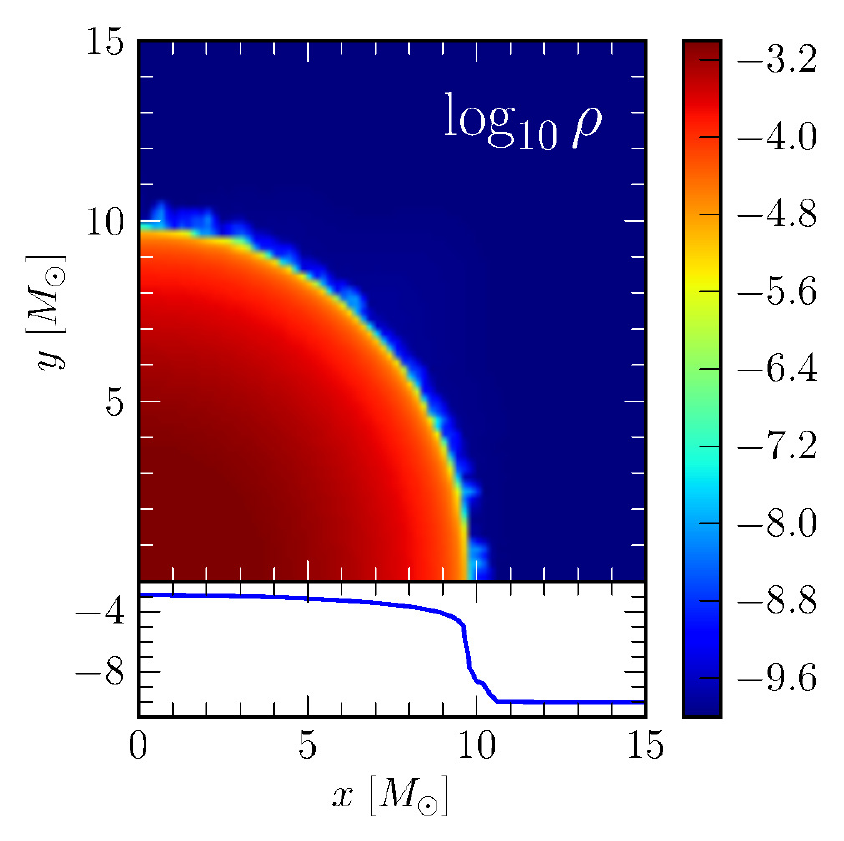}
      \\ MP5+PP
      \end{center}
  \end{minipage}
  \end{center}
  \caption{\label{fig:grthc.cowling.rho2d} Two-dimensional cut of the
    $\log_{10}$ of the rest-mass density at time $t = 800\ M_\odot$ for
    the perturbed TOV in the Cowling approximation and for different
    atmosphere prescriptions. The insets at the bottom of the plots show
    the one dimensional cuts along the $x$ axis. }
\end{figure*}

The reason for the bad behaviour of the standard MP5 prescription is
that, as anticipated in the previous Section, it lacks a sufficient
amount of numerical dissipation in the case of surfaces aligned with the
grid and especially for polytropic evolutions, such as the ones we show
here. This is clearly seen in Figure \ref{fig:grthc.cowling.rho2d} where we
show a two-dimensional cut of the rest-mass density, in $\log_{10}$
scale, at a representative time during the evolution. One can clearly see
the appearance of ``jets'' of low-density matter ($\rho \sim
10^{-5}\ M_\odot^{-2}$) aligned with the coordinate directions (see
bottom part of each panel for a one-dimensional cut along the
$x$-direction). These ``jets'' are launched at seemingly random times
from the surface of the star, when the numerical errors ``extract'' from
the atmosphere a large enough amount of rest mass. What happens is that
the numerical oscillations create an imbalance at the surface of the
star: the excess density coming from the atmosphere generates a pressure
which is only balanced by the ``potential barrier'' at the surface of the
star given by the double threshold on the rest-mass density floor. As
soon as the pressure is large enough, part of the matter is ejected in
one of these streams. This process effectively results in the increase of
the total rest mass of the star because only part of this extra matter is
actually lost from the outer boundary. In contrast, we can see that, with
the addition of extra numerical dissipation at the surface of the star,
these artefacts are completely suppressed, as shown for the MP5+LF
case. This happens partly because dissipation prevents the scheme from
extracting too much matter out of the atmosphere and partly because it
diffuses the numerical errors back into the floor. Finally, the
positivity-preserving evolution does not show any kind of numerical
ejecta out of the star's surface because of its conservative nature. On
the other hand it is affected by the accumulation of matter at the
fluid-vacuum interface. As commented before, this accumulation can be
greatly reduced by lowering the floor rest-mass density and it is also
somewhat less severe for the Gamma-law EOS case, where the floor
accretion is regulated by the thermal pressure.

\begin{figure}
\begin{center}
  \includegraphics[width=0.7\textwidth]{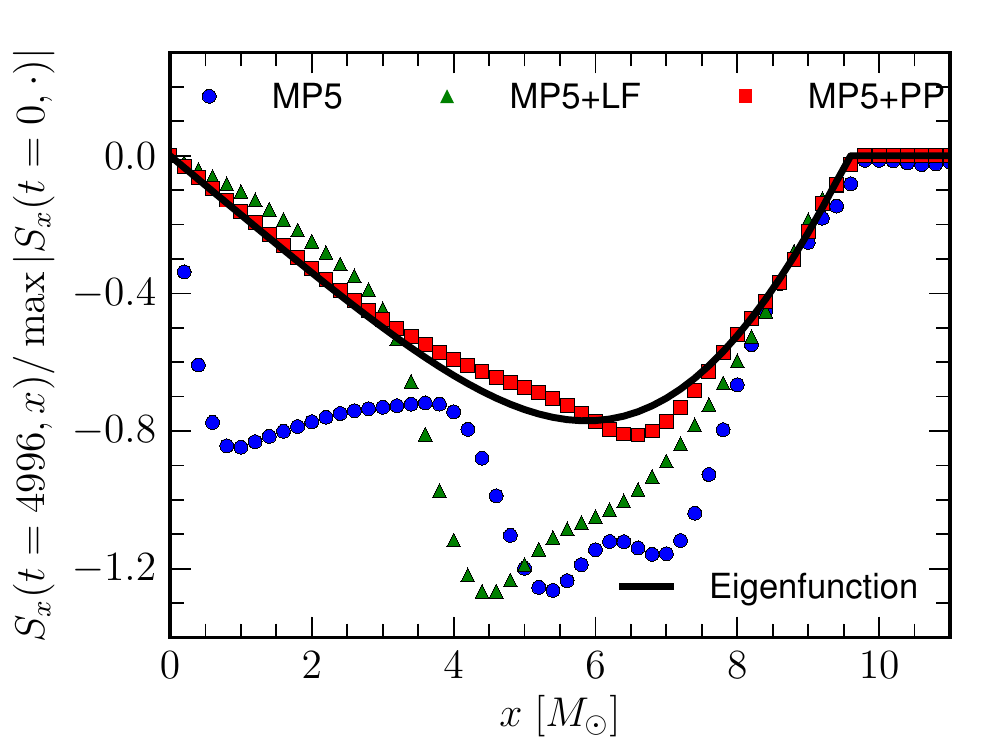}
  \caption{\label{fig:grthc.cowling.scon} One dimensional cut along the
    $x$ axis showing the $x$ component of the conserved linear momentum,
    $S_x$, at the time $t = 4996\ M_\odot$, for the perturbed TOV in the
    Cowling approximation and for different atmosphere prescriptions. The
    different symbols show the numerical solution as obtained with our
    code and with the different vacuum treatments, while the thick black
    line shows the exact eigenfunction from linear perturbation theory.
  }
\end{center}
\end{figure}

The differences between the various methods are even more evident if we
look at sensitive quantities such as the linear momentum in the radial
direction. At the initial time it has a profile given by the
eigenfunction of the $F$-mode in the Cowling approximation. In the linear
regime one would expect the linear momentum to simply oscillate with the
$F$-mode frequency. On the other hand, in a simulation, because of
numerical errors, the profile of the eigenfunction is gradually
lost. This is shown in Figure \ref{fig:grthc.cowling.scon} where we plot
the $x$-component of the linear momentum, $S_x$, along the $x$ axis at a
representative time, $t = 4996\ M_\odot$. At this particular time both
the MP5 and the MP5+LF schemes have accumulated so much error that the
profile of the eigenfunction is completely distorted. On the other hand
the evolution using the positivity-preserving limiter still shows a good
agreement with the exact solution. Clearly the precision with which we
recover the eigenfunction is resolution dependent and degrades over time
also for the MP5+PP scheme. Nevertheless, this figure clearly
illustrates: 1) how large the influence of the atmosphere is in this kind
of simulation where nearly equilibrium configurations are evolved for a
long time; 2) how small is the perturbation due to the continuous,
artificial accretion when we use our positivity preserving prescription,
even when the floor rest-mass density is rather high.

\subsection{Linear Oscillations: Full-GR}\label{sec:grthc.fullgr}

The second test we present is the evolution of a stable, nonrotating,
star in full-GR. The goal of this test is to check the stability of the
three different floor prescriptions in a fully general-relativistic
setting. The model that we consider here is the same as the one described
in Section \ref{sec:grthc.cowling}, with the difference that we do not
apply any perturbation to the initial data and we let it evolve under the
sole effects of the numerical truncation error.

\begin{table}[b]
\caption{\label{table:grthc.tov}
Numerical parameters used for the oscillating TOV test in full-GR.
}
\vspace{1em}
\begin{indented}
\item[]\begin{tabular}{ccccc}
  \br
  Model & Time integrator & $\mathrm{CFL}$ & 
    $\rho_{\mathrm{atmo}}\ [M_\odot^{-2}]$ & $\varepsilon$ \\
  \mr
  MP5    & RK4 & 0.2 & $10^{-10}$ & $1$ \\
  MP5+LF & RK4 & 0.2 & $10^{-10}$ & $0.01$ \\
  MP5+PP & RK3 & 0.2 & $10^{-19}$ & $-$ \\
  \br
\end{tabular}
\end{indented}
\end{table}

This test is performed using a grid covering $0 \leq x,y,z \leq
80\ M_\odot$ and employing three refinement levels with diameters, $20
\ M_\odot, 40\ M_\odot$ and $80\ M_\odot$, with the finest one covering
the star entirely and having a resolution $h = 0.2\ M_\odot$. The
spacetime is evolved using fourth-order finite differencing and the CCZ4
formulation of the Einstein equations. Finally, we assume reflection
symmetry across the $(x,y),\, (x,z)$ and $(y,z)$ planes. We evolve the
model with different atmosphere prescriptions and time integrators and we
choose, for each of them, the values of $\rho_{\mathrm{atmo}}$ and
$\varepsilon$ giving the best results in order to showcase the
capabilities of each method. We note, however, that, due to the high
computational costs, we did not perform an extensive tuning of these
parameters and we cannot exclude that other combinations of parameters
would give better results.  The parameters that we use are summarized in
Table \ref{table:grthc.tov}.

\begin{figure}
\begin{center}
  \includegraphics[width=0.7\textwidth]{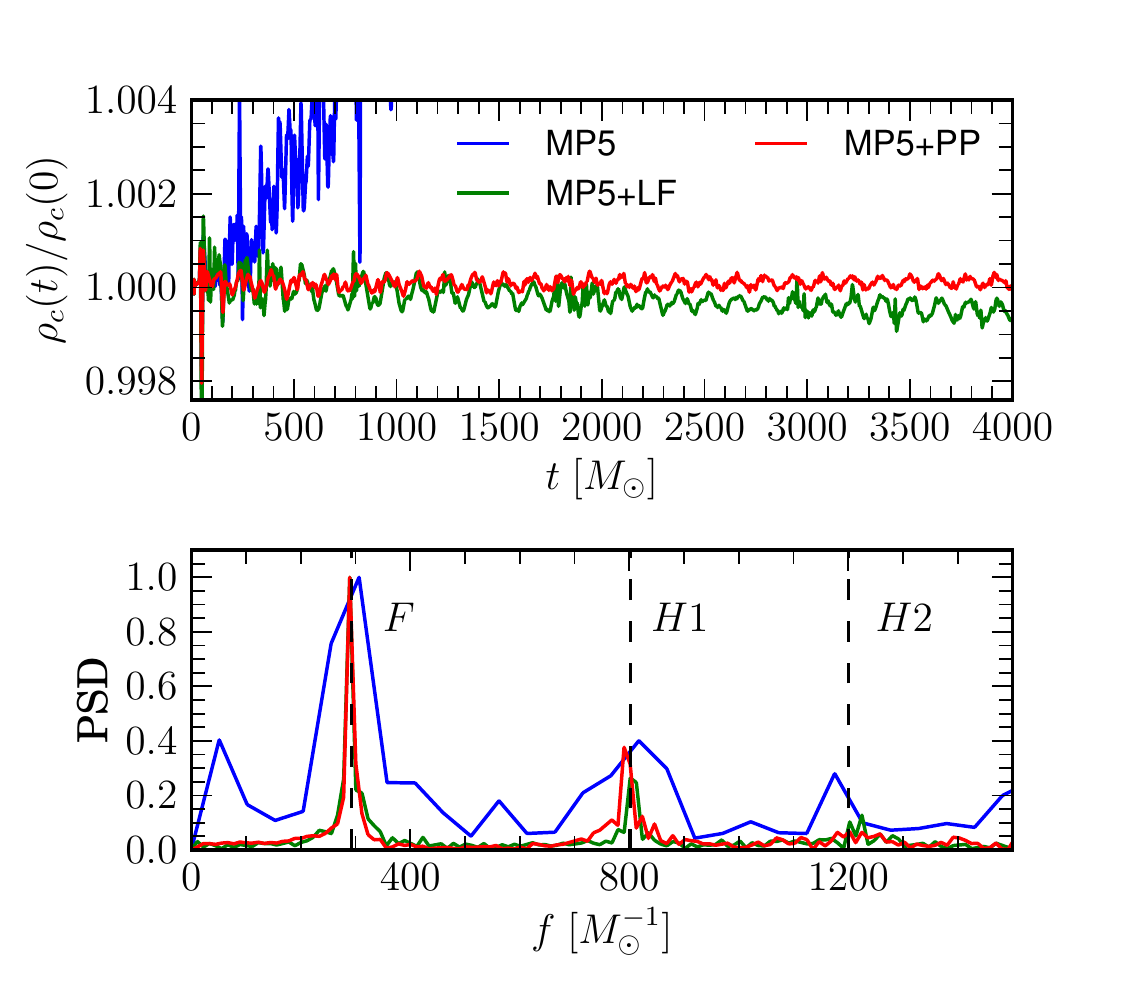}
  \caption{\label{fig:grthc.fullgr.rho} \textit{Top panel:} evolution of
    the normalized central rest-mass density for the oscillating TOV in
    full-GR and for different atmosphere prescriptions.  \textit{Bottom
      panel:} power spectral density of the central rest-mass density,
    normalized to have maximum value $1$. In the calculation of the PSD
    we exclude the first $300\ M_\odot$ of the evolution, to avoid
    contamination from the initial spike.  }
\end{center}
\end{figure}

The evolution of the central rest-mass density for the different methods
is shown in the top panel of Figure \ref{fig:grthc.fullgr.rho}. Clearly,
the ordinary MP5 prescription shows violent oscillations and a large
secular growth. We evolve this model up to time $t \simeq 1100\ M_\odot$
where it has deviations from the initial rest-mass density of the order
of $1.5\ \%$. We note that when we have evolved this model using the same
prescription as for the test in the Cowling approximation, we have
actually obtained even larger oscillations and a more pronounced secular
growth leading to an increase of about $4\ \%$ in the central rest-mass
density at time $t = 1000\ M_\odot$.

As with the previous test, the simple addition of extra numerical
dissipation at the star's surface seems to cure the most severe problems
with the MP5 evolution. Indeed the MP5+LF scheme shows much smaller
oscillations and only a weak secular trend (see also Figure
\ref{fig:grthc.cowling.rho}).

The MP5+PP scheme yields very small oscillations and an almost zero trend
in the central rest-mass density. However, in preliminary tests the
MP5+PP scheme showed a sudden increase in the oscillation amplitude and
in the secular drift after time $t \gtrsim 3000\ M_\odot$, which were at
levels comparable to the MP5+LF ones. The reason for this behaviour is to
be found in the prolongation operators used in our AMR setup as well as
in the lack of refluxing in our code, which were resulting in spurious
violations in the mass conservation at the mesh-refinement boundaries in
the low density regions outside of the star. In order to avoid this
problem we have disabled the prolongation of the hydrodynamic variables,
thus partially ``decoupling'' the various refinement levels. In the long
term we plan to improve the AMR capabilities of our code to avoid these
pathologies, but these are not of primary concern for the main purpose of
our code, which is to compute gravitational waveforms from compact
binaries.

The PSD of the central rest-mass density is shown in the bottom panel of
Figure \ref{fig:grthc.fullgr.rho}. There, we show the PSD normalized to
have maximum amplitude of $1$ for the different numerical schemes. The
spectra are computed using the central rest-mass density from the time $t
\geq 300$, to remove the dependence from the relaxation of the initial
data. In order to remove the low-frequency contribution from the secular
growth we compute the spectra after removing the secular terms via a
linear fit. Clearly, the ordinary MP5 scheme has a more noisy spectrum,
partly because of the shorter integration time. Apart from that, all the
three methods show spectra which are peaked at the frequencies
corresponding to the $F$-mode and to the first overtone, $H1$, as
computed from linear perturbation theory.

\begin{figure}
\begin{center}
  \includegraphics[width=0.7\textwidth]{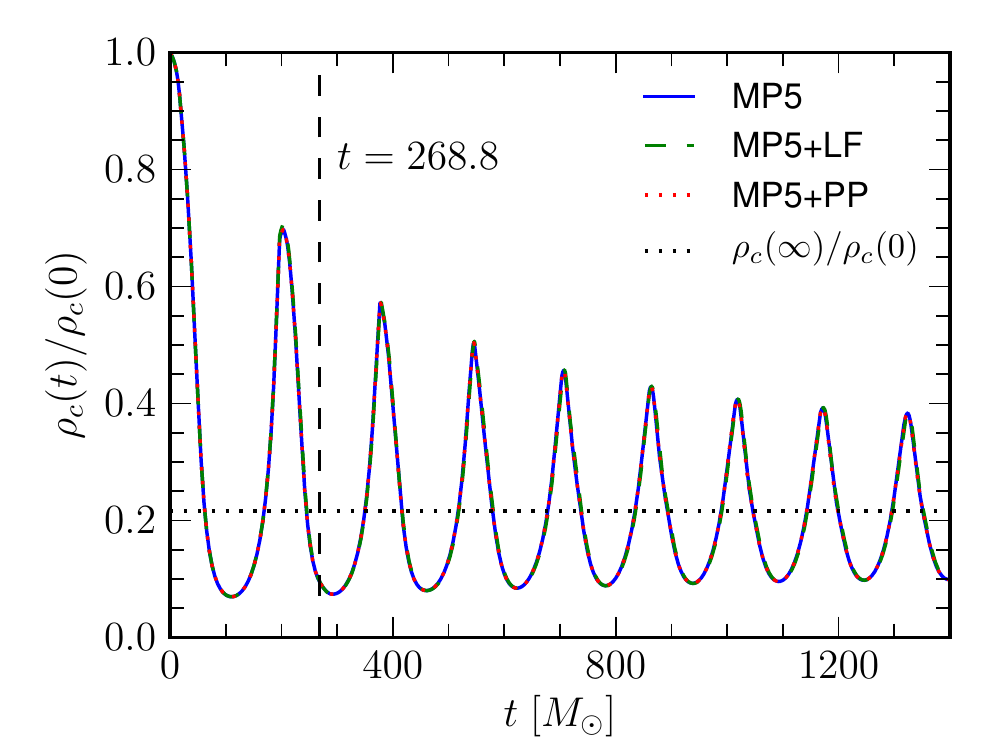}
  \caption{\label{fig:grthc.migration.rho} Evolution of the central
    rest-mass density, normalized to its initial value, for the TOV
    migration test and for different numerical schemes.  The vertical
    dashed line marks the point in time shown in
    Figure \ref{fig:grthc.migration.rho2d}. The horizontal dotted line
    marks the value of the central rest-mass density for the equilibrium
    model on the stable branch corresponding to the unstable model
    evolved in this test and to which the solution is expected to relax.
  }
\end{center}
\end{figure}

\subsection{Nonlinear Oscillations: the Migration Test}

The third test that we discuss is the study of the large, nonlinear,
oscillations of a TOV migrating from the unstable branch of equilibrium
solutions to the stable one. This is a commonly adopted test for
numerical-relativity codes, \eg \cite{Font02c, Baiotti03a, Baiotti04,
  Cordero2009, Thierfelder2011}, and has been studied in detail by
\cite{liebling_2010_emr, Radice:10}. Here we consider a model initially
described by a polytropic EOS with $\Gamma=2$ and $K=100$ and with
central rest-mass density $\rho_c = 0.007\ M_\odot^{-1}$, yielding an ADM
mass of $\simeq 1.49\ M_\odot$. The migration is triggered by the use of
an outgoing velocity perturbation of the form $v^r = A\ r$, where $r$ is
the areal radius and $A$ is chosen so that the maximum perturbation
velocity is $0.01$. The evolution is performed with a Gamma-law equation
of state to allow for shock heating. We do not solve the constraints
equations after the application of the initial perturbation, but we rely
on the constraint-damping nature of CCZ4 to bring the evolution back to
the constraint ``hypersurface'' as done in \cite{Kastaun2013}.

The grid setup is identical to the one described in Section
\ref{sec:grthc.fullgr}. Here, again, we use the same atmosphere
prescription for all the schemes with $\rho_{\mathrm{atmo}} =
10^{-12}\ M_\odot^{-2}$ and we evolve all the models using the SSP RK3
scheme with a CFL of $0.1$. Finally, the spacetime is evolved using
sixth-order finite differencing and with the addition of a fifth-order
Kreiss-Oliger dissipation.

The evolution of the system is summarized by
Figure \ref{fig:grthc.migration.rho}, where we show the evolution of the
central rest-mass density, normalized to its initial value, for our three
different schemes. As can be seen from the figure, the star undergoes a
sequence of violent expansion, contraction cycles after it has migrated
on the stable branch of equilibria, while conserving the baryonic
mass. During the contraction phase, shocks are formed and part of the
shock-heated matter is ejected with large velocities from the central
object. All of the methods are perfectly adequate for this test and only
minimal differences appear between the MP5+LF scheme and the other two in
the amplitude of the first peak.

\begin{figure*}
  \begin{minipage}{0.5\hsize}
    \begin{center}
      \includegraphics[width=1.0\textwidth]{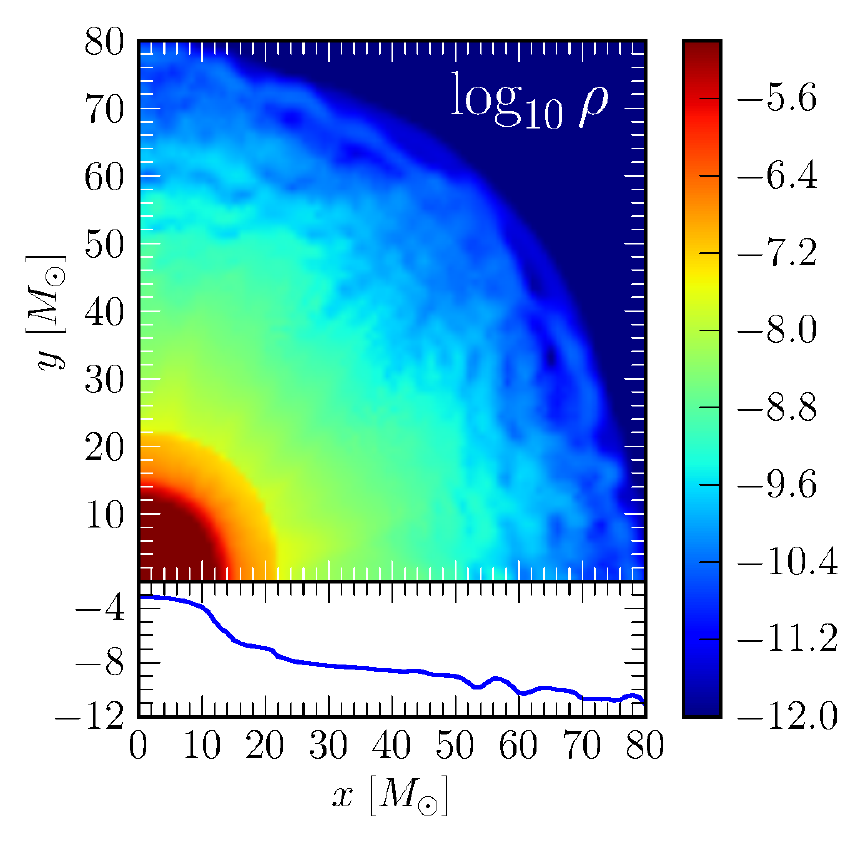}
      \\ MP5
    \end{center}
  \end{minipage}
  \begin{minipage}{0.5\hsize}
    \begin{center}
      \includegraphics[width=1.0\textwidth]{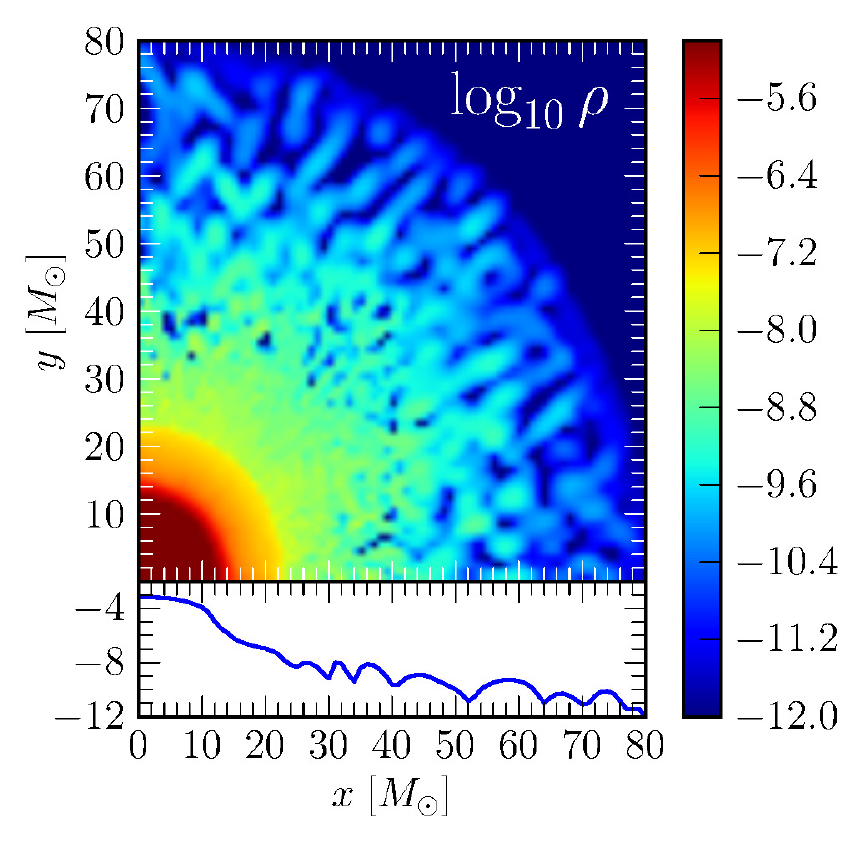}
      \\ MP5+LF
    \end{center}
  \end{minipage}
  \begin{center}
  \begin{minipage}{0.5\hsize}
    \begin{center}
      \includegraphics[width=1.0\textwidth]{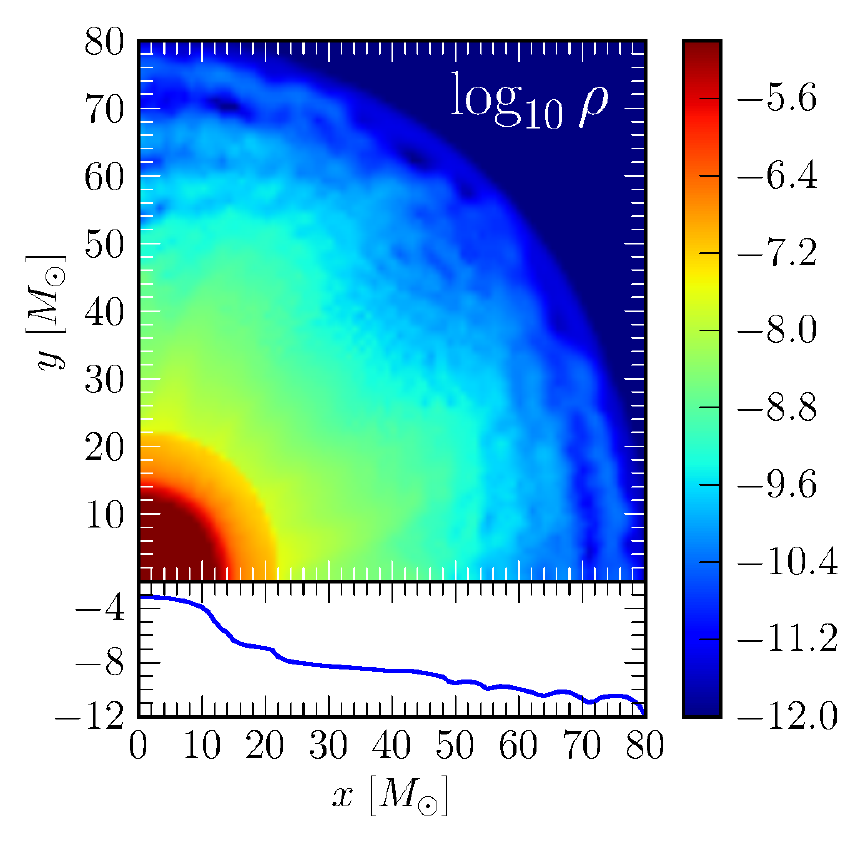}
      \\ MP5+PP
    \end{center}
  \end{minipage}
  \end{center}
  \caption{\label{fig:grthc.migration.rho2d} Two-dimensional cuts of the
    $\log_{10}$ of the rest-mass density for the TOV migration test and
    for different numerical schemes at time $t = 268.8\ M_\odot$. The
    insets show the one-dimensional cuts along the $x$ axis. }
\end{figure*}

The difference between the various atmosphere prescriptions is better
appreciated by looking at Figure \ref{fig:grthc.migration.rho2d}, where we
show a two-dimensional cut of the $\log_{10}$ of the rest-mass density at
the time when the matter ejected at the first bounce reaches the grid
boundaries (this is indicated with a vertical line in
Figure \ref{fig:grthc.migration.rho}). As can be seen from the figure, both
MP5 and MP5+PP are able to capture the dynamics of the low-density ejecta
without introducing large numerical oscillations or excessive deviations
from spherical symmetry. The front of the ejecta is reasonably well
captured even if it has crossed two mesh-refinement boundaries, where our
code cannot currently ensure mass-conservation and hence the right
propagation speed for shocks. On the other hand, the MP5+LF method
exhibits large numerical oscillations, which lead to small islands of
larger rest-mass density. This is probably due to our choice of avoiding
the reconstruction in characteristic variables at low densities, since it
is well known that component-by-component reconstruction typically
results in oscillatory solutions when used in conjunction with high-order
schemes. Of course, given the smallness of the rest mass which is
concentrated in these fragments, the bulk dynamics of the matter is
essentially unaltered and the spectral properties of the oscillating star
are as in the other methods.

\subsection{Gravitational Collapse to Black-Hole}
\begin{figure}
\begin{center}
  \includegraphics[width=0.7\textwidth]{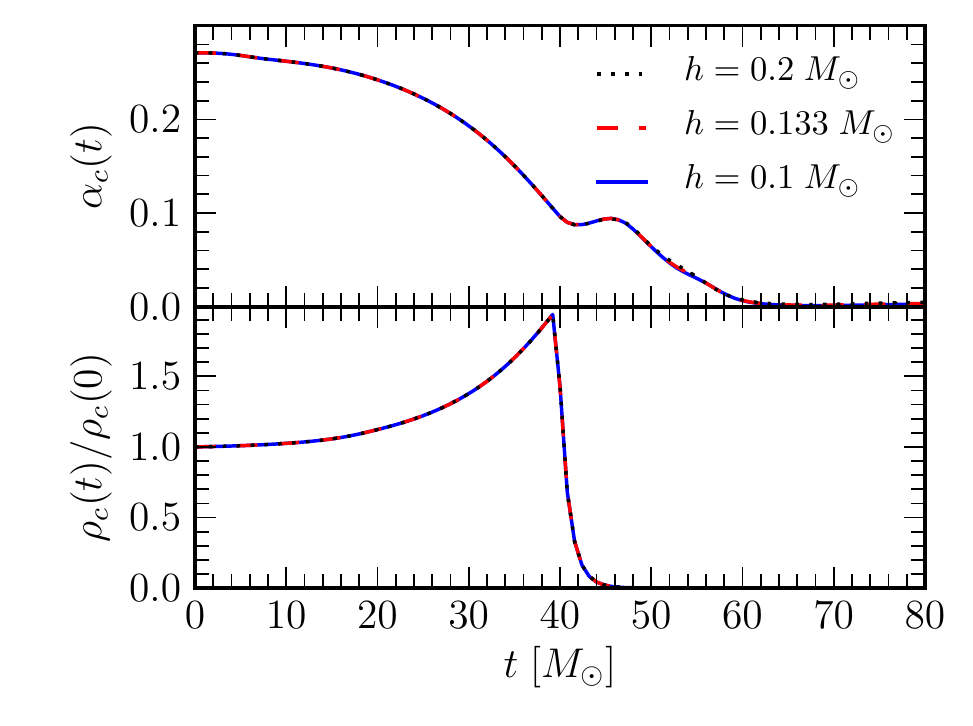}
  \caption{\label{fig:grthc.collapse.alprho} \textit{Top panel:}
    evolution of the normalized central lapse for the TOV collapse test
    and for different resolutions. \textit{Bottom panel:} evolution of
    the normalized central rest-mass density for the TOV collapse test
    and for different resolutions. }
\end{center}
\end{figure}

The final test involving isolated neutron stars that we describe is the
gravitational collapse of a TOV to a black hole. This is another commonly
adopted benchmark for general-relativistic hydrodynamics codes and has
been studied in great detail in
\cite{Baiotti06,Baiotti07,Thierfelder10}. The model that we consider here
is initially described by a polytropic EOS with $\Gamma = 2$ and $K =
100$ and has an initial central rest-mass density of
$0.008\ M_\odot^{-2}$, yielding an ADM mass of $\simeq
1.43\ M_\odot$. The collapse is triggered with the addition of a velocity
perturbation with the same properties of the one employed in the
migration test, but with opposite sign, \ie an ingoing perturbation. The
model is evolved using an Gamma-law EOS to allow for thermal effects.

In the case of the collapse the influence of the atmosphere is
negligible, so we consider only evolutions performed with the MP5+LF
prescription. As in the migration test, our computational domain covers
$0 \leq x,y,z \leq 80\ M_\odot$ and we assume reflection symmetry across
the $(x,y),\, (x,z)$ and $(y,z)$ planes. We employ four different
refinement levels with the finest one covering the star entirely with
diameters $10\ M_\odot, 20\ M_\odot, 40\ M_\odot$ and $80\ M_\odot$ we
study the convergence of the code as we vary the resolution. In
particular we considered six different resolutions having grid spacing
(in the finest refinement level) of $h = 0.2, 0.16, 0.13333, 0.11429,
0.1$ respectively.  We also perform a higher-resolution run, with
$h=0.08$, which we evolve only up to time $\simeq 45\ M_\odot$ and that
we use as a reference solution to measure the self-convergence of the
code. We adopt sixth-order finite differencing for the spacetime, which
is evolved using the CCZ4 formulation, with fifth-order Kreiss-Oliger
artificial dissipation on the metric variables. We use the fourth-order
Runge-Kutta scheme as time integrator, so that our scheme is formally
fourth-order (fifth-order in space and fourth in time). Finally, in order
to avoid excessive oscillations in the matter fields inside the forming
black hole, we artificially evacuate the regions where $\alpha < 0.1$ by
adding an artificial damping term in the sources of the hydrodynamic
variables as done in \cite{Alic2013}. We also switch to the
component-wise reconstruction with Lax-Friedrichs split in regions where
$\alpha < 0.2$.

The evolution of the lapse and the rest-mass density at the coordinate
origin and for the different resolutions are shown in
Figure \ref{fig:grthc.collapse.alprho}. As the star collapses the central
rest-mass density rapidly increases and the lapse function approaches
zero. At time $t\simeq 40\ M_\odot$ the lapse at the center becomes
smaller than $0.1$ and the rest-mass density starts to be
dissipated. Finally, after a small re-bounce, the lapse settles and the
evolution reaches quasi-stationarity. The exact behaviour of the lapse is
determined by the way in which we evacuate the fluid, since we found that
in simulations without matter damping, the lapse showed a smaller bounce
at the moment of the collapse, but a more irregular evolution at
intermediate times.

\begin{figure}
\begin{center}
  \includegraphics[width=0.7\textwidth]{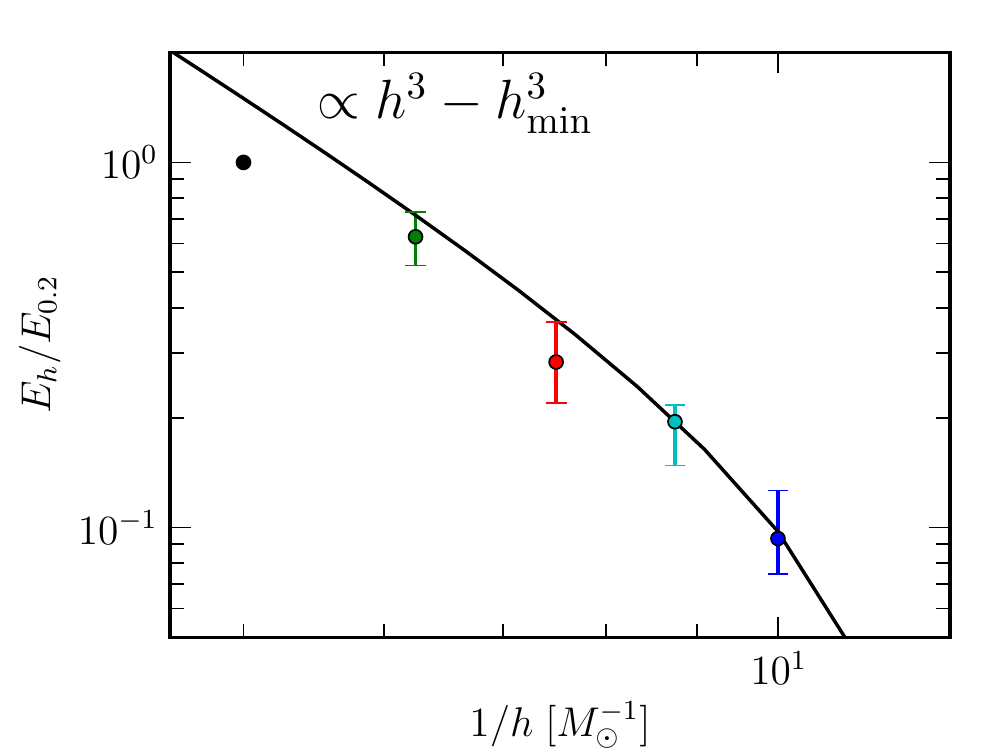}
  \caption{\label{fig:grthc.collapse.alp_convergence} Estimated
    $L^1-$norm of the error of the lapse function on the $(x,y)$ plane on
    the finest refinement level for the TOV collapse test and for
    different resolutions. The errorbars show the excursion between the
    maximum and minimum normalized error in the time interval. The solid
    black line shows the curve for third-order of convergence.  }
\end{center}
\end{figure}

In order to estimate the convergence rate of our code we use the highest
resolution simulation, $h = 0.08$, as a reference solution and we compute
the error of a given physical quantity, $\phi$, at the time $t_n$ as
\begin{equation}
  E_h^n = \frac{1}{N}\sum_{x}| \phi_h(x,t_n) - \phi_{h=0.08}(x,t_n) |\,,
\end{equation}
where the sum is taken over the common grid points between the resolution
$h$ and the highest resolution run on the $(x,y)$ plane. $E_h^n$ is then
computed using $14$ data-sets equally spaced in time the interval
$t/M_{\odot} [3.2, 44.8]$ (including the first and the last time). In
order to have an absolute measure of the relative errors between the
different resolutions over the whole time-interval, we normalize the
error estimates with respect to the deviations as measured between the
lowest resolution simulation and the reference one, \ie
$E_h^n/E_{0.2}^n$.  Finally, we take as relative error the average in time
of the normalized error estimates and we use the maximum and minimum
relative errors (between the different times) as a measure of the
uncertainty of this procedure.

The results obtained for the lapse function are shown in
Figure \ref{fig:grthc.collapse.alp_convergence}. Also shown, as a solid
black line, is the curve for third-order convergence. As can be seen from
the figure, our data is consistent with third-order convergence for $h
\gtrsim 0.16$. As commented before, our code is formally fourth-order
convergent in time and fifth-order in space, on the other hand, based on
our previous experience with MP5 in \cite{Radice2012a}, we argue
that the observed third-order convergence is most probably related to the
fact that high-order shock-capturing codes are able to converge at their
nominal order only at extremely high resolutions. This is due to the fact
that their nominal accuracy is typically spoiled by the activation of the
flattening procedure close to under resolved features of the solution
(see discussion in \cite{Radice2012a}).

\section{Binary Neutron Stars}
\label{sec:grthc.bns}

In this Section we present results obtained for the inspiral and merger
of binary neutron stars in quasi-circular orbit. We consider a binary
having an initial small separation of $45\ \mathrm{km}$, as this binary
can be simulated with relatively small computational costs, allowing us
to explore the different atmosphere prescriptions and make a detailed
comparison between the results obtained with our code and the ones
obtained with the original \texttt{Whisky} code. The same binary, but at
the larger separation of $60\ \mathrm{km}$, has been considered in
\cite{Radice2013b}.

We recall here that the \texttt{Whisky} code is a second-order
finite-volume code with high-order primitive reconstruction and
implements several different approximate Riemann solvers. For the runs
presented here we make use of the PPM reconstruction \cite{Colella84} and
of the HLLE Riemann solver \cite{Harten83, Einfeldt88}. We remark that
\texttt{Whisky} is a good representative of the current state-of-the-art
for numerical general-relativistic hydrodynamics \cite{Baiotti:2010ka}.

The initial data is computed in the conformally-flat approximation using
the \texttt{\textsc{LORENE}} pseudo-spectral code \cite{Gourgoulhon01}
and is publicly available \cite{lorene41}. The EOS assumed for the initial
data is polytropic with $K = 123.56$ and $\Gamma=2$, while the evolution
is performed using the Gamma-law EOS to allow for thermal effects in the
merger phase. The details of the binaries we consider are listed in Table
\ref{table:grthc.bns}, but it is important to point out that the neutron
stars composing these binaries have a rather high baryonic mass, $M_b
\simeq 1.9\ M_\odot$, close to the maximum mass allowed by the EOS for
nonrotating models, $M_{b,\max} = 2\ M_\odot$, and having high
compactness, $\mathcal{C} = M_\infty/R_\infty = 0.18002$, $M_\infty$
being the gravitational mass of each of the two stars when at infinite
separation and $R_\infty$ the corresponding areal radius. Binaries with
a similar compactness have been already considered in 
\cite{Hotokezaka2013b}, where it was found that high-compactness binaries
are much more challenging to evolve accurately with respect to
low-compactness ones.

\begin{table}
\caption{\label{table:grthc.bns} Summary of the neutron-stars binaries
  considered. For each binary, we report the total baryonic mass, $M_b$,
  the ADM mass, $M_{_{\mathrm{ADM}}}$, the initial separation, $r$ and
  the initial orbital frequency $f_{\mathrm{orbit}}$, the gravitational
  mass of each of the two stars when at infinite separation, $M_\infty$,
  as well as the compactness, $\mathcal{C} = M_\infty/R_\infty$, where
  $R_\infty$ is the areal radius of the two stars when at infinite
  separation. }
\vspace{1em}
\begin{indented}
\item[]\begin{tabular}{ccccccc}
  \br
  Binary &
    $M_b\ [M_\odot]$ &
    $M_{_{\mathrm{ADM}}}\ [M_\odot]$ &
    $r\ [\mathrm{km}]$ &
    $f_{\mathrm{orb}}\ [\mathrm{Hz}]$ &
    $M_\infty\ [M_\odot]$ &
    $\mathcal{C}$ \\
  \mr
  A & $3.8017$ & $3.44537$ & $45$ & $309.702$ & $1.7428$ & $0.18002$ \\
  B & $3.8017$ & $3.45366$ & $60$ & $208.431$ & $1.7428$ & $0.18002$ \\
  \br
\end{tabular}
\end{indented}
\end{table}
\begin{table}
\caption{\label{table:grthc.runs}
Summary of the main numerical parameters used in the numerical
simulations presented here. For each run we give the name of the code
used to perform it, \texttt{WhiskyTHC} or \texttt{Whisky}, the numerical
method employed, the time integrator used for the method of lines, MOL,
the CFL, the number of refinement levels of the grid,
$N_{\mathrm{refl}}$, and the grid spacing in the finest refinement level,
$h$.
}
\vspace{1em}
\begin{indented}
\item[]\begin{tabular}{lllllll}
  \br
  Run           & Code               & Method & MOL & CFL &
  $N_{\mathrm{refl}}$ & $h\ [M_\odot]$ \\
  \mr
  \texttt{A.MP5.H1}      & \texttt{WhiskyTHC} & MP5    & RK4 & 0.30 & 5 & 0.40000 \\
  \texttt{A.MP5.H2}      & \texttt{WhiskyTHC} & MP5    & RK4 & 0.30 & 6 & 0.20000 \\
  \texttt{A.MP5.H4}      & \texttt{WhiskyTHC} & MP5    & RK4 & 0.30 & 6 & 0.12800 \\
  \texttt{A.MP5+LF.H2}   & \texttt{WhiskyTHC} & MP5+LF & RK4 & 0.30 & 6 & 0.20000 \\
  \texttt{A.MP5+PP.H2}   & \texttt{WhiskyTHC} & MP5+PP & RK3 & 0.15 & 6 & 0.20000 \\
  \texttt{A.PPM.H2}      & \texttt{Whisky}    & PPM    & RK4 & 0.30 & 6 & 0.20000\\
  \texttt{A.PPM.H3}      & \texttt{Whisky}    & PPM    & RK4 & 0.30 & 6 & 0.13333 \\
  \texttt{A.PPM.H5}      & \texttt{Whisky}    & PPM    & RK4 & 0.30 & 6 & 0.10000 \\
  \br
\end{tabular}
\end{indented}
\end{table}

In what follows we discuss the results obtained from eight different
evolutions of the binary A described in Table \ref{table:grthc.bns}. As a
comparison, in the table we also present the properties of the binary we
have considered in \cite{Radice2013b}, \ie binary B. All of these runs
are performed on a grid covering $0 < x,z \leq 512\ M_\odot$,
$-512\ M_\odot \leq y \leq 512\ M_\odot$, where we assume reflection
symmetry across the $(x,y)$ plane and $\pi$ symmetry across the $(y,z)$
plane. The grid employs several \emph{fixed} refinement levels, $5$ or
$6$ depending on the run, with the finest refinement levels covering both
stars, \ie we have no moving boxes. The refinement levels have diameters
on the equatorial plane of $30, 40, 64, 120, 240$ and $512\ M_\odot$ (the
finest one is removed for run \texttt{A.MP5.H1}). A summary of the main
numerical parameters can be found in Table \ref{table:grthc.runs}.

Finally, we evolve this model using the CCZ4 formulation with damping
constant $\kappa_1 = 0.036$ and with beta-driver $\eta = 0.71$. The
spacetime is evolved using fourth-order finite differencing and with
fifth-order Kreiss-Oliger artificial dissipation. The evolutions are
performed \textit{without} resetting the shift to zero at the beginning
of the simulation, which is known to yield a more oscillatory behaviour
in the coordinates \cite{Baiotti08}. There is no particular reason for
this choice: the gauges are only chosen so as to be able to leverage, in
the debugging stage, on the comparison with previously existing
\texttt{Whisky} simulations that were performed, with a different grid
setup, by \cite{Alic2013}.

Since our focus here is mostly on the accuracy of the calculation of the
gravitational radiation from compact neutron-star binaries, we consider
the accuracy of the code by mainly looking at the $\ell = 2, m = 2$ mode
of the Weyl scalar $\Psi_4$ extracted at the fixed coordinate radius of
$r = 450\ M_\odot$ ($\simeq 130\ M_{_{\mathrm{ADM}}}$). We do not attempt
to extrapolate $\Psi_4$ in radius or compute the strain as this involves
other uncertainties \cite{Boyle:2009vi, Reisswig:2009us, Reisswig:2010a,
  Reisswig:2011}.

\begin{figure*}
  \begin{minipage}{0.5\hsize}
    \begin{center}
      \includegraphics[width=1.0\hsize]{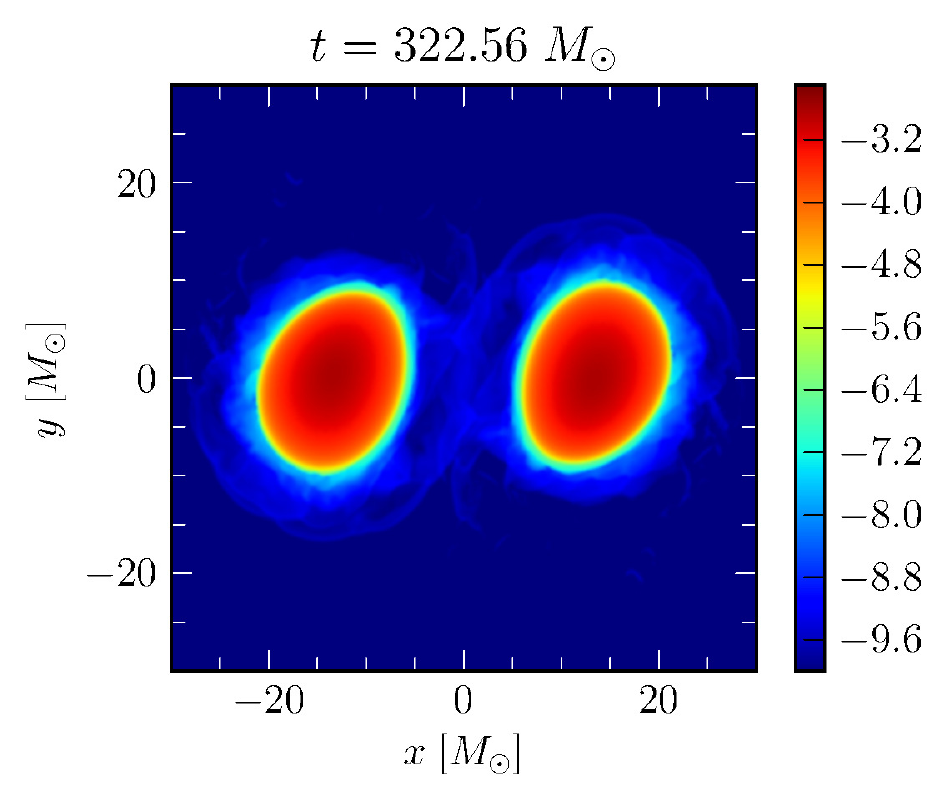}\\
      \includegraphics[width=1.0\hsize]{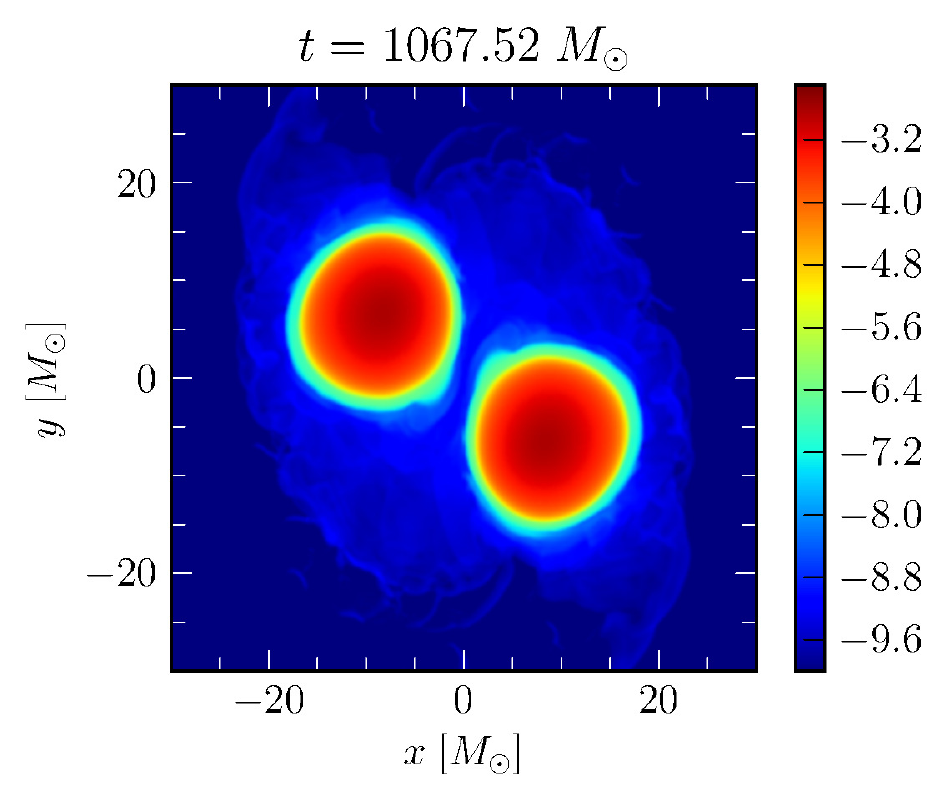}\\
      \includegraphics[width=1.0\hsize]{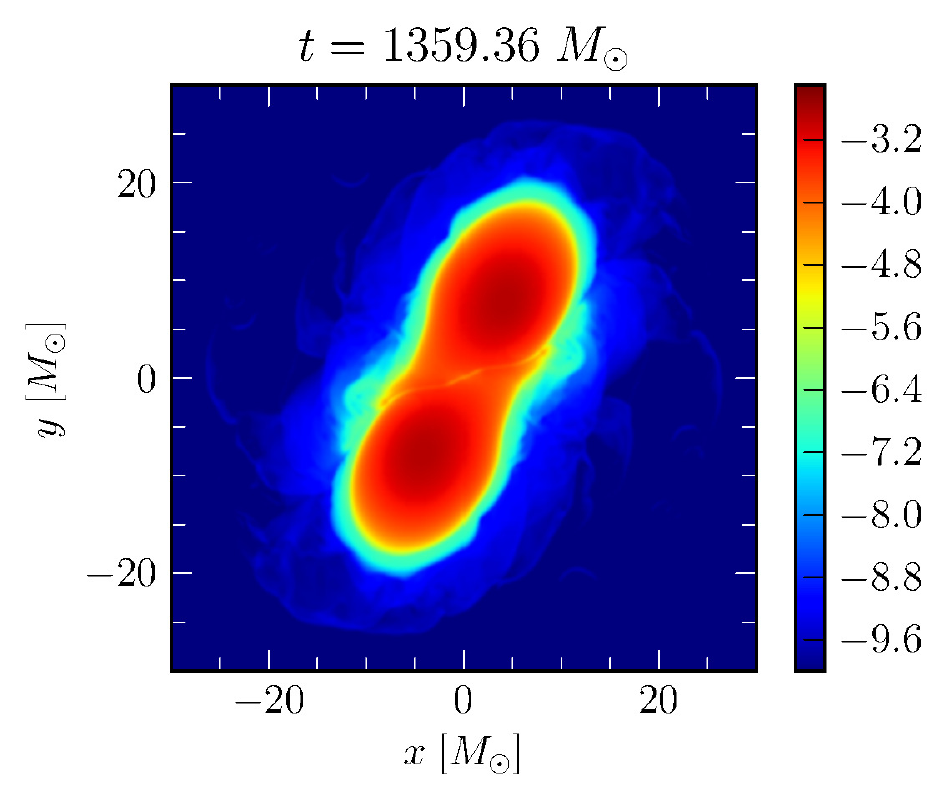}
    \end{center}
  \end{minipage}
  \begin{minipage}{0.5\hsize}
    \begin{center}
      \includegraphics[width=1.0\hsize]{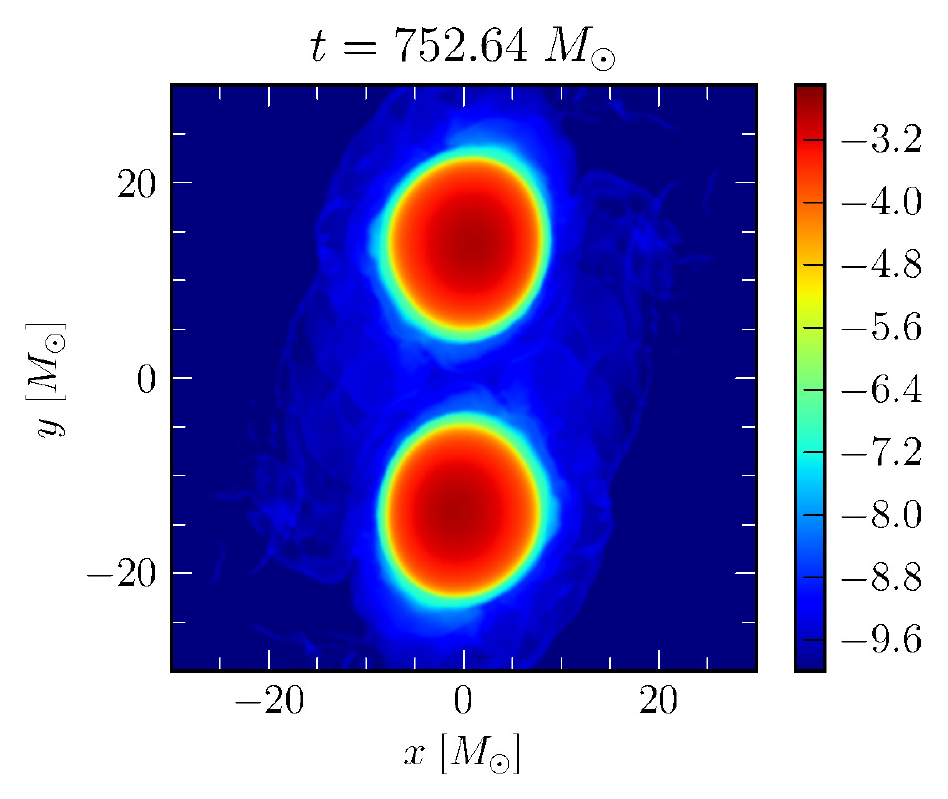}\\
      \includegraphics[width=1.0\hsize]{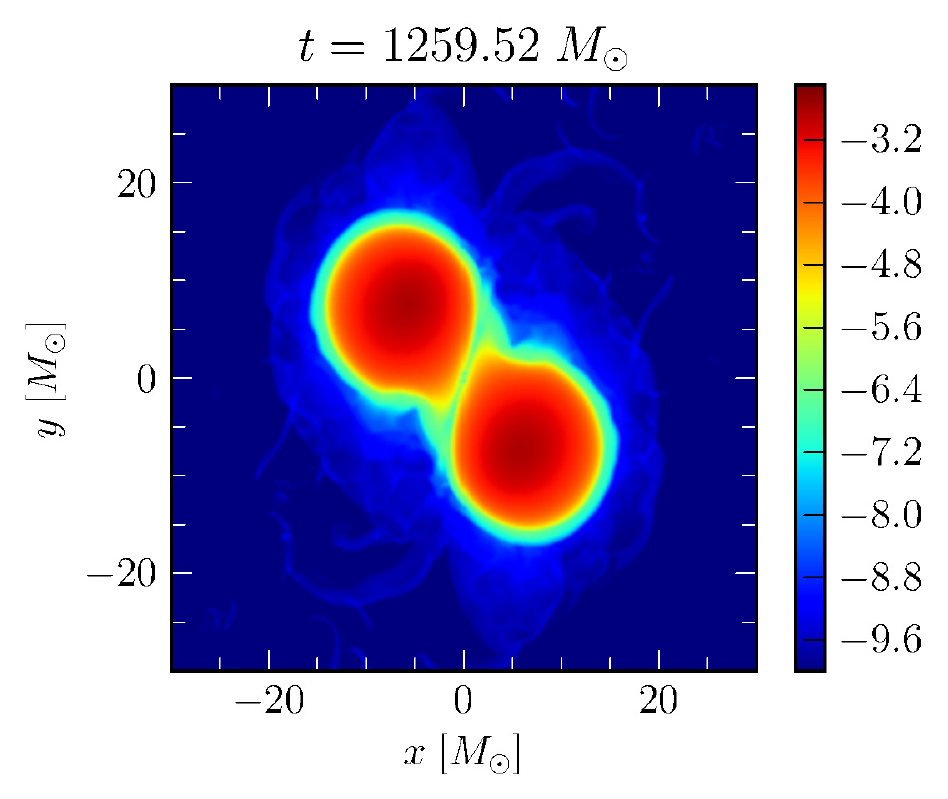}\\
      \includegraphics[width=1.0\hsize]{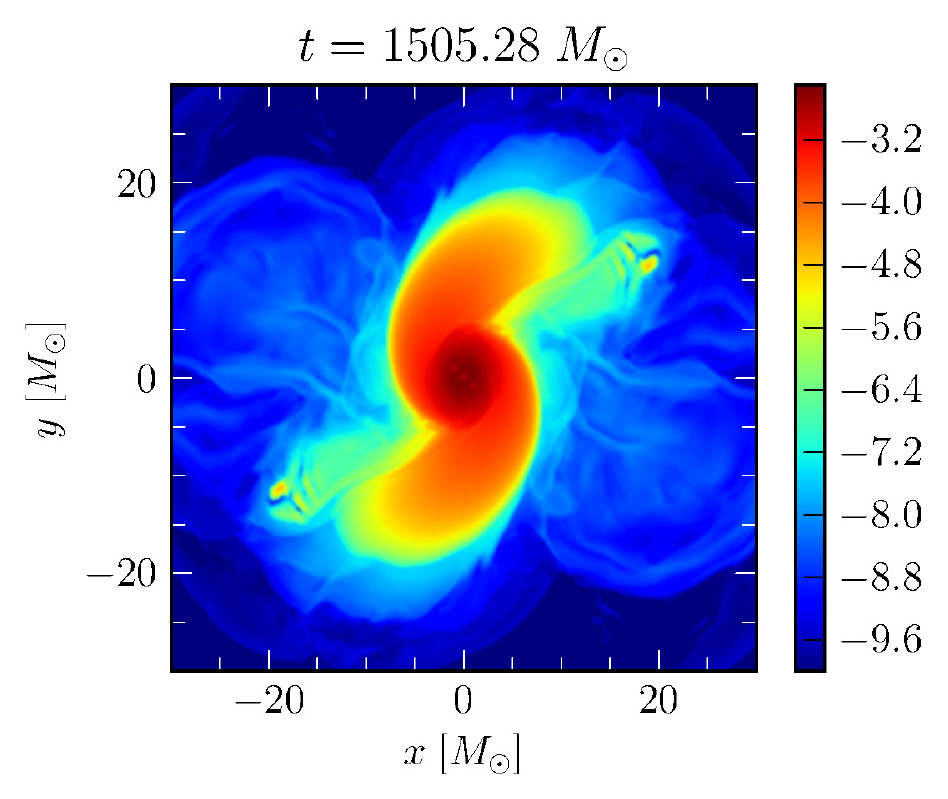}
    \end{center}
  \end{minipage}
  \caption{\label{fig:bns60.thc.rho2d} Two-dimensional visualization of
    the $\log_{10}$ of the rest-mass density for the run
    \texttt{A.MP5.H2}. The results have been obtained with the
    \texttt{WhiskyTHC} code; see main text for details.}
\end{figure*}

\begin{figure*}
  \begin{minipage}{0.5\hsize}
    \begin{center}
      \includegraphics[width=1.0\hsize]{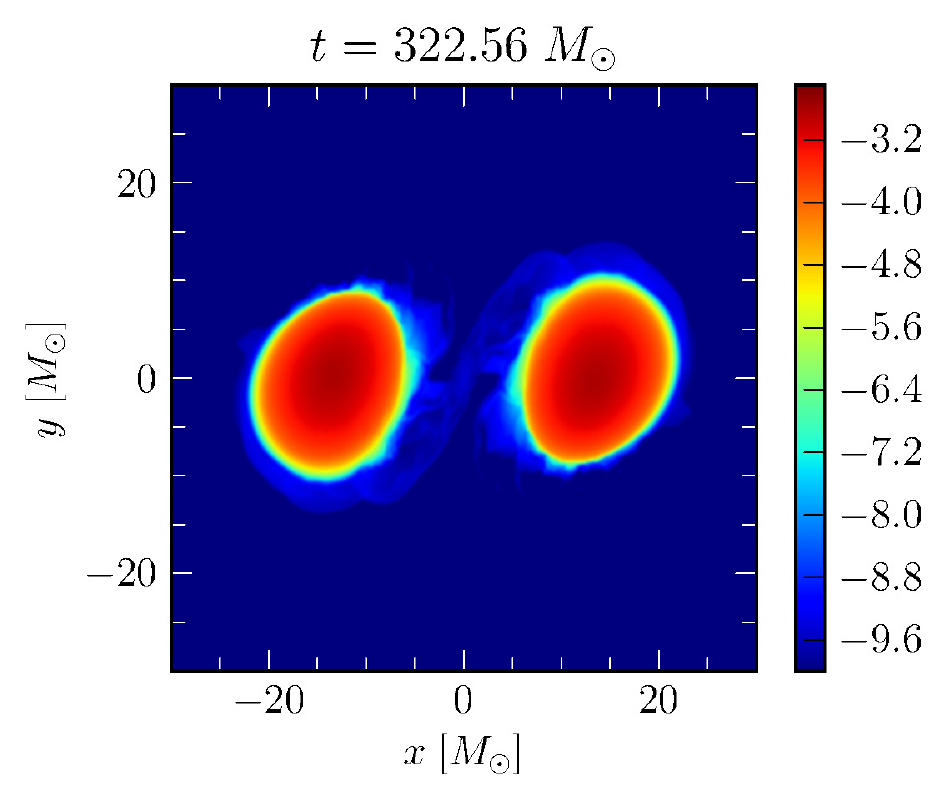}\\
      \includegraphics[width=1.0\hsize]{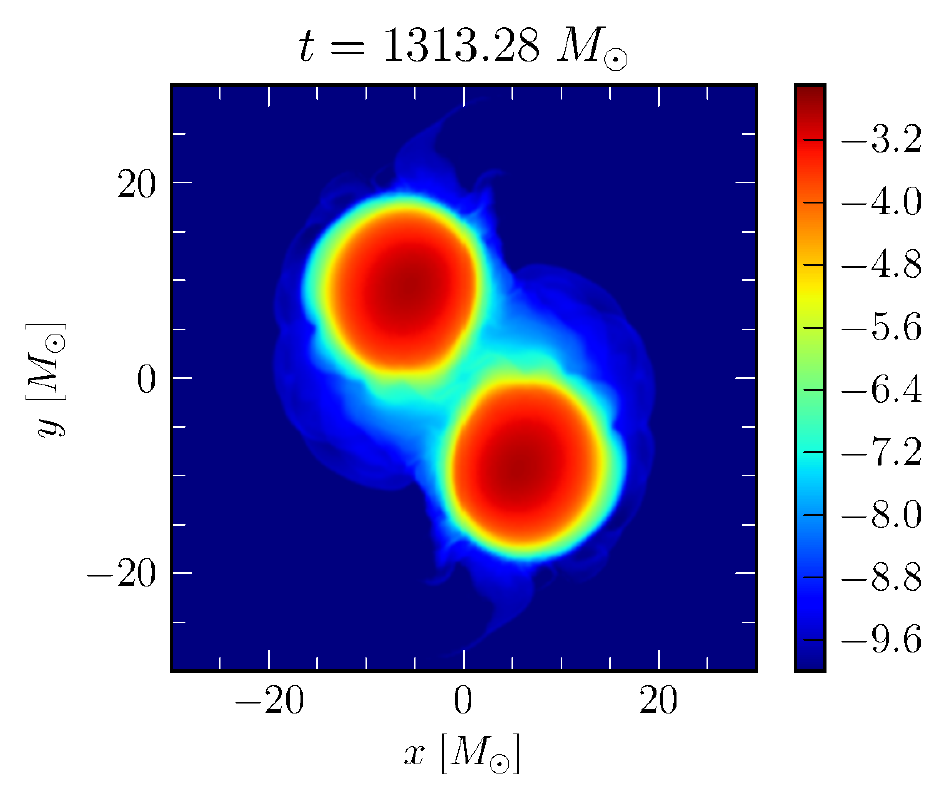}\\
      \includegraphics[width=1.0\hsize]{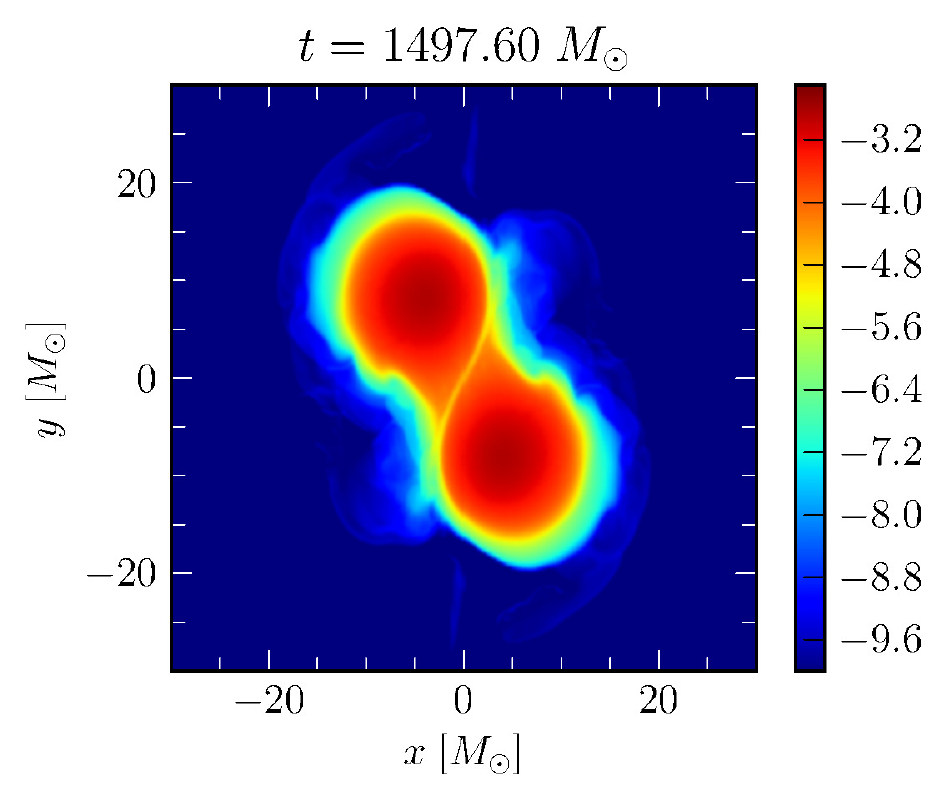}
    \end{center}
  \end{minipage}
  \begin{minipage}{0.5\hsize}
    \begin{center}
      \includegraphics[width=1.0\hsize]{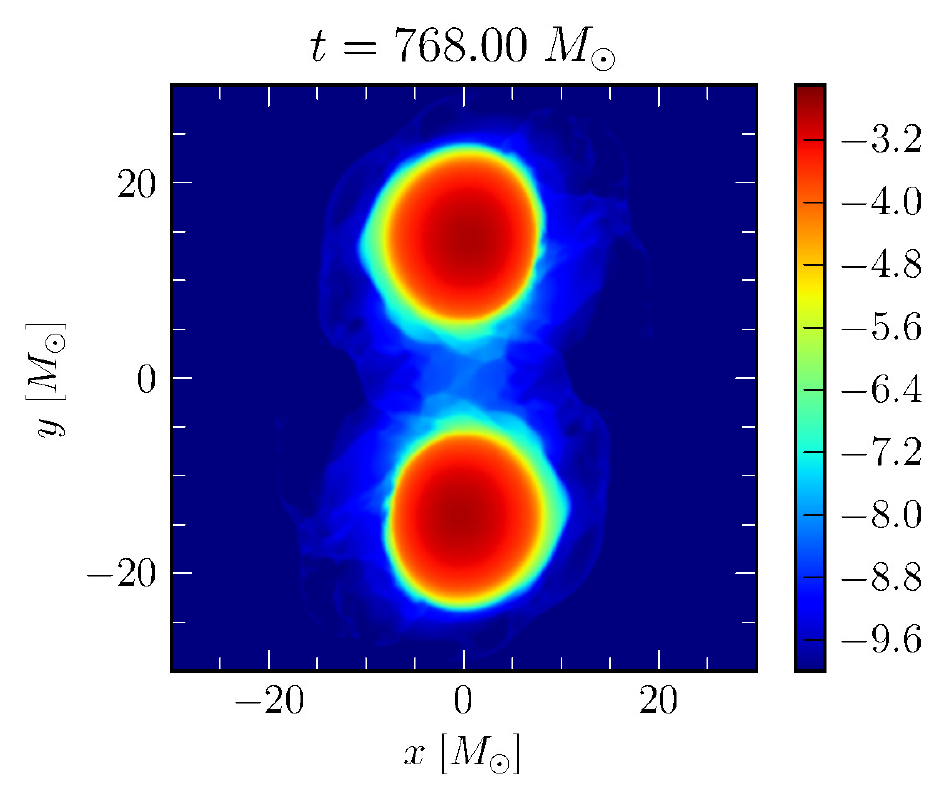}\\
      \includegraphics[width=1.0\hsize]{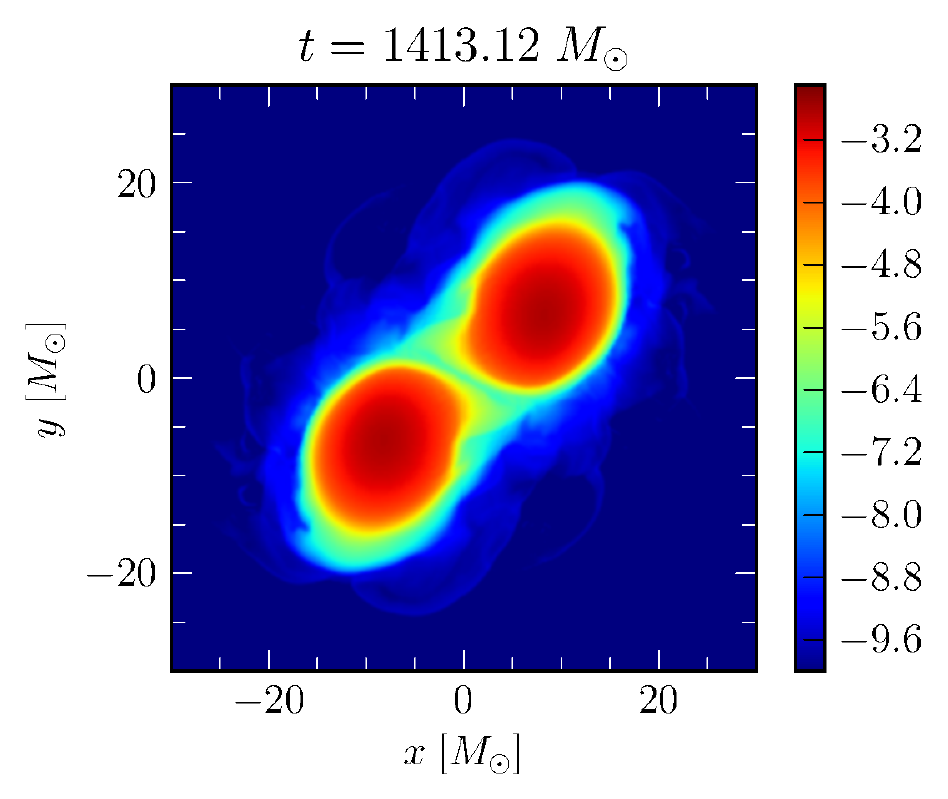}\\
      \includegraphics[width=1.0\hsize]{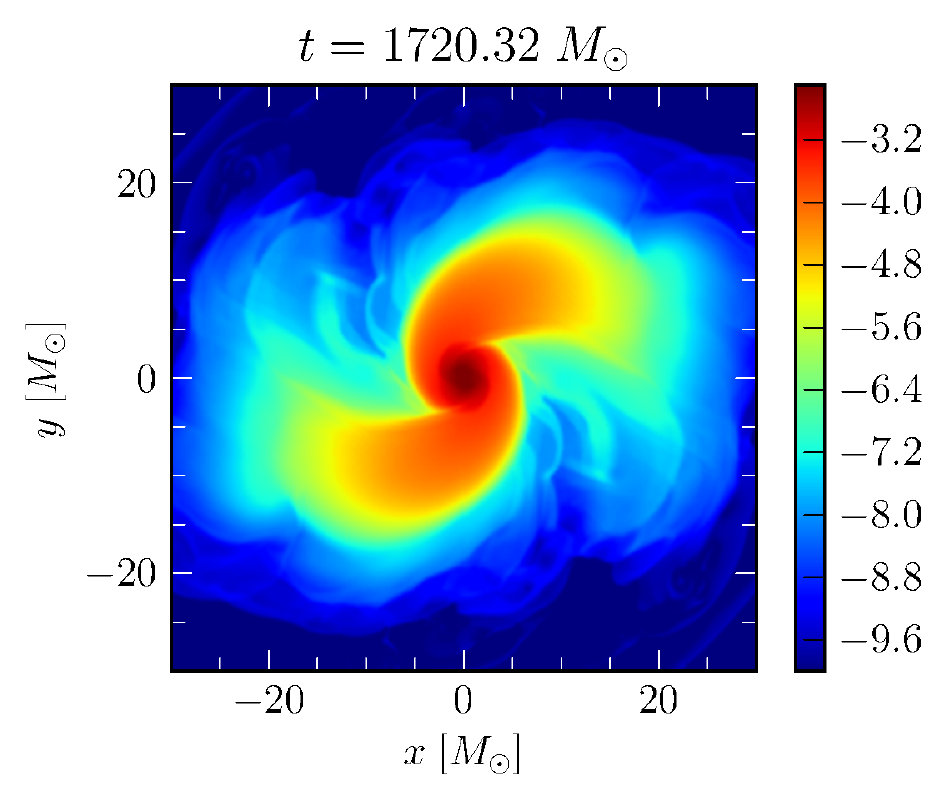}
    \end{center}
  \end{minipage}
  \caption{\label{fig:bns60.whisky.rho2d} Two-dimensional visualization
    of the $\log_{10}$ of the rest-mass density for the run
    \texttt{A.PPM.H2}. The results have been obtained with the
    \texttt{Whisky} code; see main text for details. }
\end{figure*}

The dynamics of the inspiral and merger of binary neutron stars has been
described many times and in great detail in the literature, \eg
\cite{Baiotti08}; for this reason we do not give a very in-depth
discussion of it here. The two neutron stars inspiral for about $2.5$
orbits, touch and quickly merge into a black-hole. For this particular
model no significant disk is left behind. The gravitational-wave signal
consists of about seven cycles up to merger, followed by the black-hole
ringdown.

An overview of the inspiral and merger dynamics is shown in Figures
\ref{fig:bns60.thc.rho2d} and \ref{fig:bns60.whisky.rho2d} for the
\texttt{WhiskyTHC} and \texttt{Whisky} codes, respectively, at the common
resolution of $h=0.2\ M_\odot$. There we plot the $\log_{10}$ of the
rest-mass density $\rho$ on the $(x,y)$ plane at six representative
times.  As remarked in \cite{Baiotti08}, the large deformations of
the stars is only an apparent one and is due mostly to the large
deformation of the coordinates arising from the use of an initially
non-zero shift as specified by the \texttt{LORENE} initial data.

From a simple comparison of Figures \ref{fig:bns60.thc.rho2d} and
\ref{fig:bns60.whisky.rho2d}, it is possible to appreciate that
\texttt{WhiskyTHC} is able to preserve the initial sharp profile of the
two neutron stars for the whole duration of the inspiral and up to the
merger. In comparison, the second-order methods of the original
\texttt{Whisky} code result in a significant smearing of the rest-mass
density profile. As a consequence, the contact appears to be more smooth
in the simulations made with \texttt{Whisky} than with \texttt{WhiskyTHC}
and, in particular, with the latter one it is possible to notice the
formation of strong shock waves at the moment of the contact that eject,
but do not unbind, part of the neutron-star matter in the direction of
the separatrix between the two stars (\cf last panel of Figure
\ref{fig:bns60.thc.rho2d}). This is consistent with what described by the
toy model proposed in \cite{Kyutoku2012} in the case of equal-mass
binaries.

\begin{figure}
\begin{center}
  \includegraphics[width=0.7\textwidth]{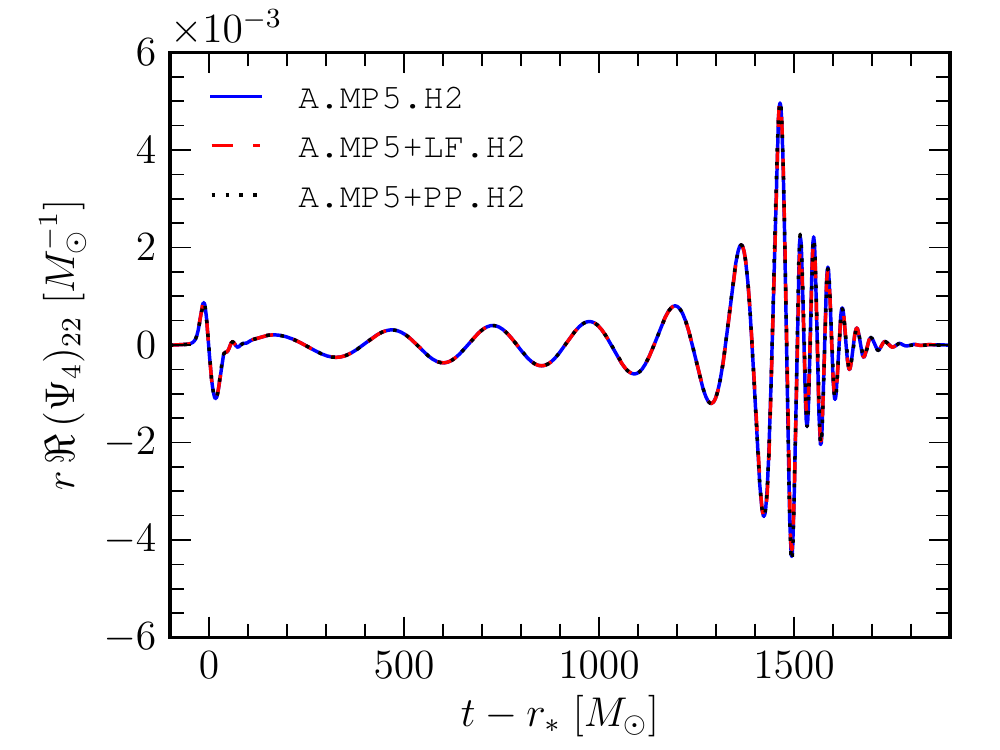}
  \caption{\label{fig:grthc.bns45.datmo.rpsi4} Real part of the $\ell=2,
    m=2$ mode of $\Psi_4$ extracted at $r = 450\ M_\odot$ for model A and
    for different atmosphere prescriptions using \texttt{WhiskyTHC}.  }
\end{center}
\end{figure}

The gravitational-wave signal is shown in Figure
\ref{fig:grthc.bns45.datmo.rpsi4}, where we plot the $\ell=2, m=2$ mode
of $\Psi_4$, as extracted at $r = 450\ M_\odot$, and as a function of the
retarded time $t - r_*$, where $r_* \equiv r + 2 M_{_{\mathrm{ADM}}} \log
[ r/(2 M_{_{\mathrm{ADM}}}) - 1]$. In particular, we show the results
obtained for runs \texttt{A.MP5.H2}, \texttt{A.MP5+LF.H2} and
\texttt{A.MP5+PP.H2}, where obviously the three names refer to
simulations performed with the MP5, the MP5+LF, and MP5+PP methods,
respectively.

As can be seen from the plot, all our three different atmosphere
prescriptions give identical results during the inspiral and yield very
marginal differences in the merger phase. This provides an important
result and suggests that the treatment of the neutron star surface is not
a leading source of error in binary-neutron-star simulations, as far as
the inspiral gravitational-wave signal is concerned. The particular
choice of time-integrator, between SSP-RK3 and the standard RK4, also
seems not to be of fundamental importance here, with the error being most
likely dominated by the spatial discretization. On the other hand, in
order to use the proper positivity-preserving limiter, the timestep used
for the \texttt{A.MP5+PP.H2} run is only half of the one used in the
other runs thus possibly introducing a systematic difference.

\begin{figure*}
  \begin{minipage}{0.5\hsize}
    \begin{center}
      \includegraphics[width=\textwidth]{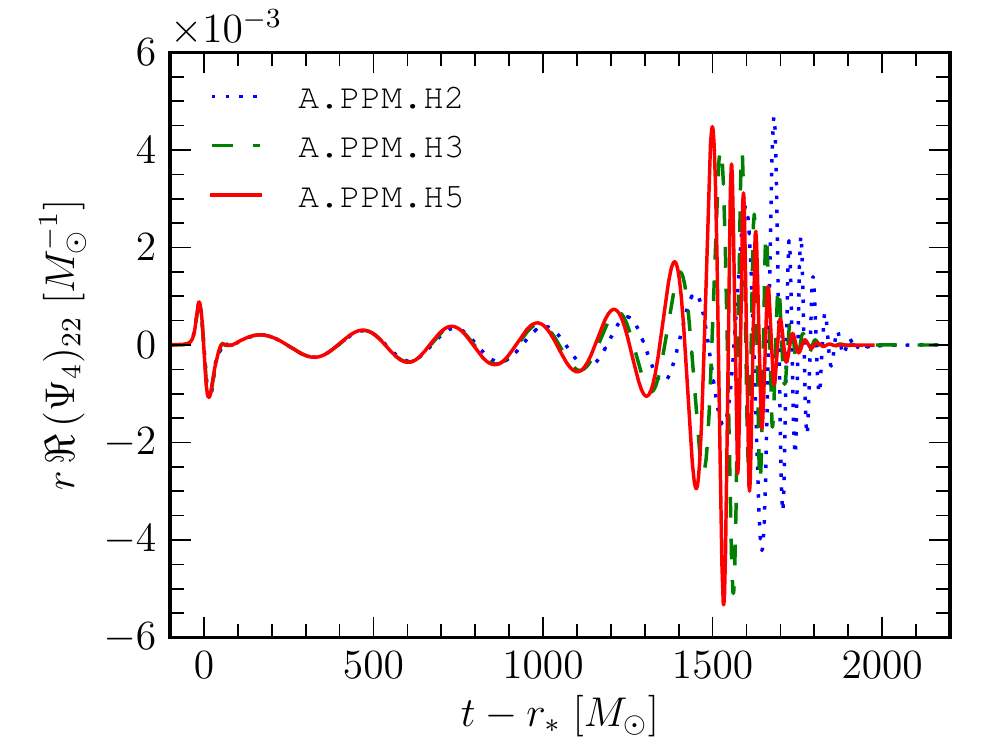} \\
      \texttt{Whisky}
    \end{center}
  \end{minipage}
  \hspace{0.05\hsize}
  \begin{minipage}{0.5\hsize}
    \begin{center}
      \includegraphics[width=\textwidth]{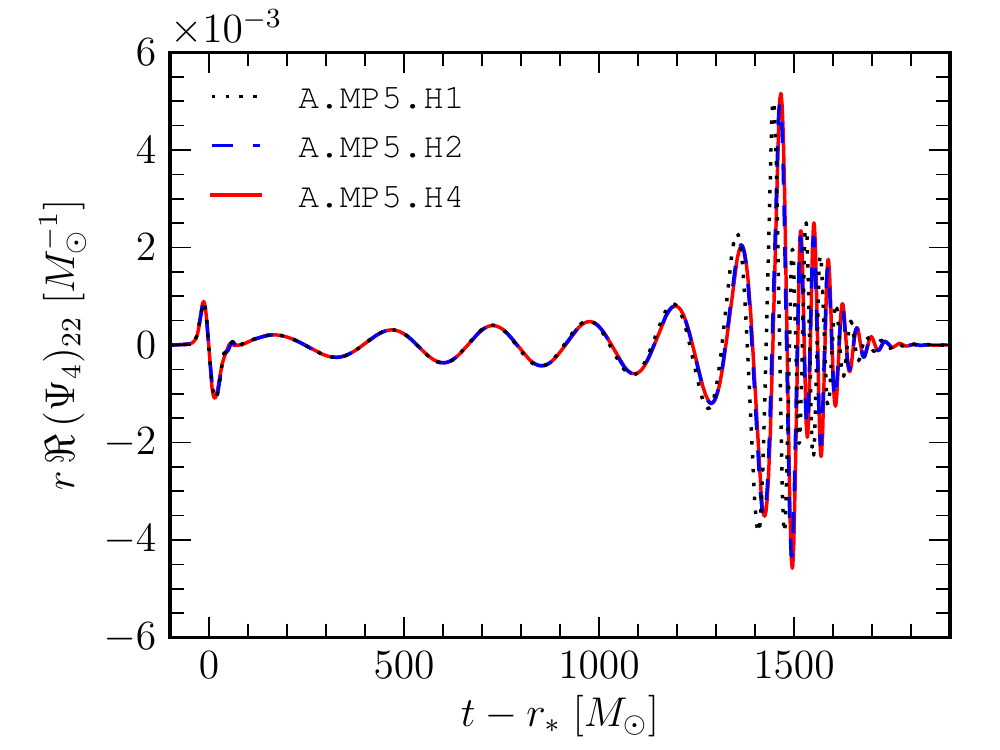} \\
      \texttt{WhiskyTHC}
    \end{center}
  \end{minipage}
  \caption{\label{fig:bns45.rpsi4} Real part of the $\ell=2, m=2$ mode of
    $\Psi_4$ extracted at $r = 450\ M_\odot$ for binary A at different
    resolutions and using two different codes: the original
    \texttt{Whisky} code and the new \texttt{WhiskyTHC} code.  }
\end{figure*}

The gravitational-wave signal for the other runs of binary A are shown in
Figure \ref{fig:bns45.rpsi4}, where in the left panel we show the results
obtained with the standard \texttt{Whisky} code and in the right one
those obtained with \texttt{WhiskyTHC}. The first aspect to notice when
comparing the two panels is that, when using a second-order code, the
phase difference between the gravitational-waves at different resolutions
is significant. We can observe a difference between the low and the high
resolution of about $\simeq 2$ radians at $t - r_* = 1350\ M_\odot$, with
$r=450\ M_\odot$ being the extraction radius. In contrast, the waveforms
obtained with \texttt{WhiskyTHC} show a significantly smaller de-phasing:
the difference between the low and the high resolution is about $\simeq
0.6$ radians at $t - r_* = 1350\ M_\odot$, which is a factor four smaller
than the one shown by \texttt{Whisky}, even though the \texttt{WhiskyTHC}
runs span a wider range of resolutions. The difference in phase between
the high and the medium-resolution simulation of \texttt{WhiskyTHC} at $t
- r_* = 1350\ M_\odot$ is as small as $\simeq 0.06$ radians. The second
interesting aspect is that, for this particular binary and with the
\texttt{Whisky} code, the merger takes place \textit{earlier} as we
increase the resolution. This is the opposite of what it is observed in
other, less compact binaries, \eg \cite{Bernuzzi2012}, where tidal
effects have been found to be amplified at lower resolution, or is shown
by the \texttt{WhiskyTHC} code.

\begin{figure*}
  \begin{minipage}{0.5\hsize}
    \begin{center}
      \includegraphics[width=\textwidth]{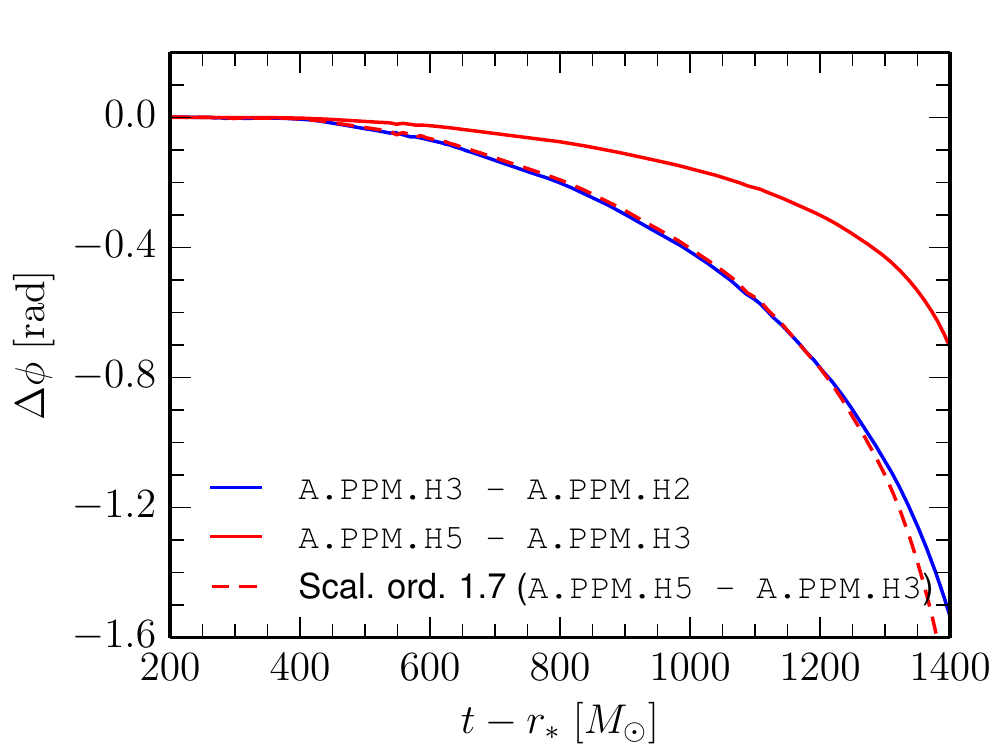} \\
      \texttt{Whisky}
    \end{center}
  \end{minipage}
  \hspace{0.05\hsize}
  \begin{minipage}{0.5\hsize}
    \begin{center}
      \includegraphics[width=\textwidth]{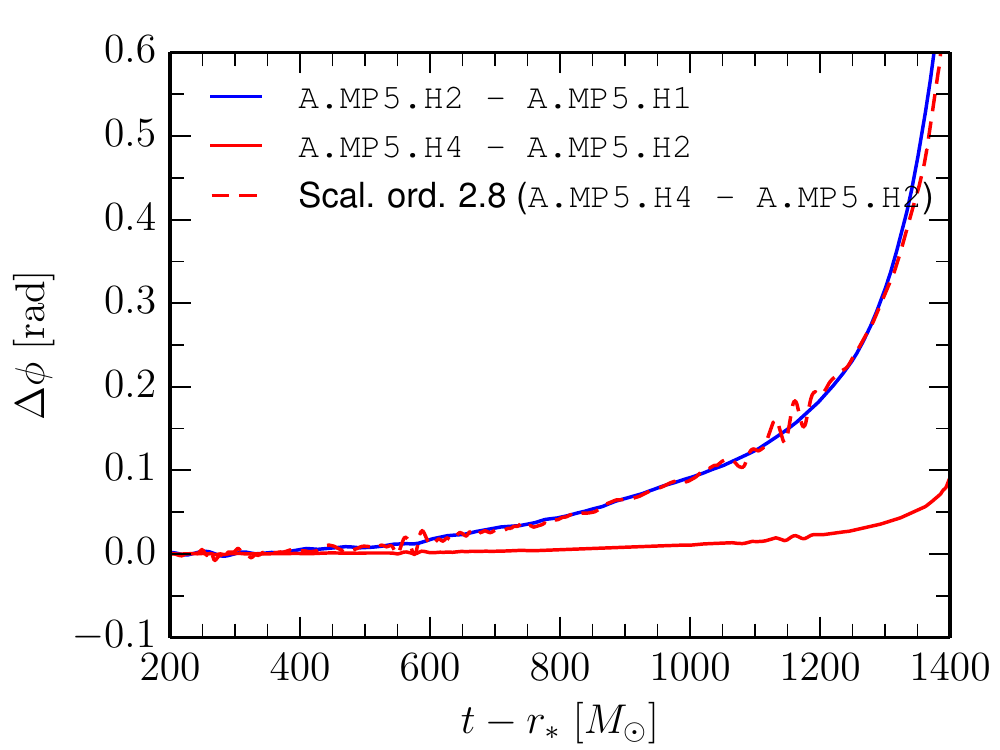} \\
      \texttt{WhiskyTHC}
    \end{center}
  \end{minipage}
  \caption{\label{fig:bns45.phase}
  Accumulated de-phasing between different resolutions for binary A as
  evolved with \texttt{Whisky} and \texttt{WhiskyTHC}. In both plots we
  show the de-phasing between the medium and low resolution (blue lines),
  between the high and medium resolution (green lines) as well as the
  rescaled de-phasing between high and medium resolution (red lines)
  computed assuming convergence order of 1.7 and 2.8 for \texttt{Whisky}
  and \texttt{WhiskyTHC} respectively.
  }
\end{figure*}

We study the convergence of the waveforms by looking at the de-phasing
between different resolutions. For each run we compute the phase, $\phi$,
of the $\ell=2, m=2$ mode of $\Psi_4$ from its definition,
\begin{equation}
  (\Psi_4)_{22} = A\ e^{i \phi},
\end{equation}
over the time interval $t - r_* \in [200, 1400]\ M_\odot$. Note that we
\textit{do not} align the waveforms at a given time, as done instead in
alignment of the waves from different resolutions \cite{Baiotti2011,
Hotokezaka2013b}, and which is hard to justify from a mathematical point
of view. On the other hand, we exclude from the calculation the first
burst of gravitational radiation, due to the initial ``junk'' radiation
present in the initial data. We also exclude the last part of the merger
phase (where we expect large errors due to the presence of shock waves)
and ringdown, since we are only concerned with the inspiral phase here.

The results are shown in Figure \ref{fig:bns45.phase}, where we measure the
convergence rate for both the \texttt{Whisky} and the \texttt{WhiskyTHC}
codes. In particular, for \texttt{Whisky} (\cf left panel of Figure
\ref{fig:bns45.phase}) we find a convergence order of $\simeq 1.7$
essentially up to time $t - r_* \simeq 1200\ M_\odot$, \ie essentially up
to the contact time $t\simeq 1200\ M_\odot$; this convergence order is
comparable with the one reported in \cite{Baiotti:2009gk} and which
was $\simeq 1.8$. For \texttt{WhiskyTHC}, instead (\cf right panel of
Figure  \ref{fig:bns45.phase}), we find a convergence order of $\simeq 2.8$
up to time $t - r_* \simeq 1300\ M_\odot$. It is useful to remark that
together with the results reported in \cite{Radice2013b}, where a
similar convergence order of $\simeq 3.2$ was found, this is the first
time that higher-than-second-order of convergence has been shown for
binary-neutron-star mergers\footnote{We note that there are several
  reasons that could explain why the convergence order found here is
  slightly smaller than the one found in \cite{Radice2013b}. The
  most important among these are that we are considering shorter
  inspirals here and the CCZ4 formulation that normally requires larger
  resolutions to enter the convergence
  regime~\cite{Alic:2011a,Alic2013}.}.

\begin{figure}
\begin{center}
  \includegraphics[width=0.7\textwidth]{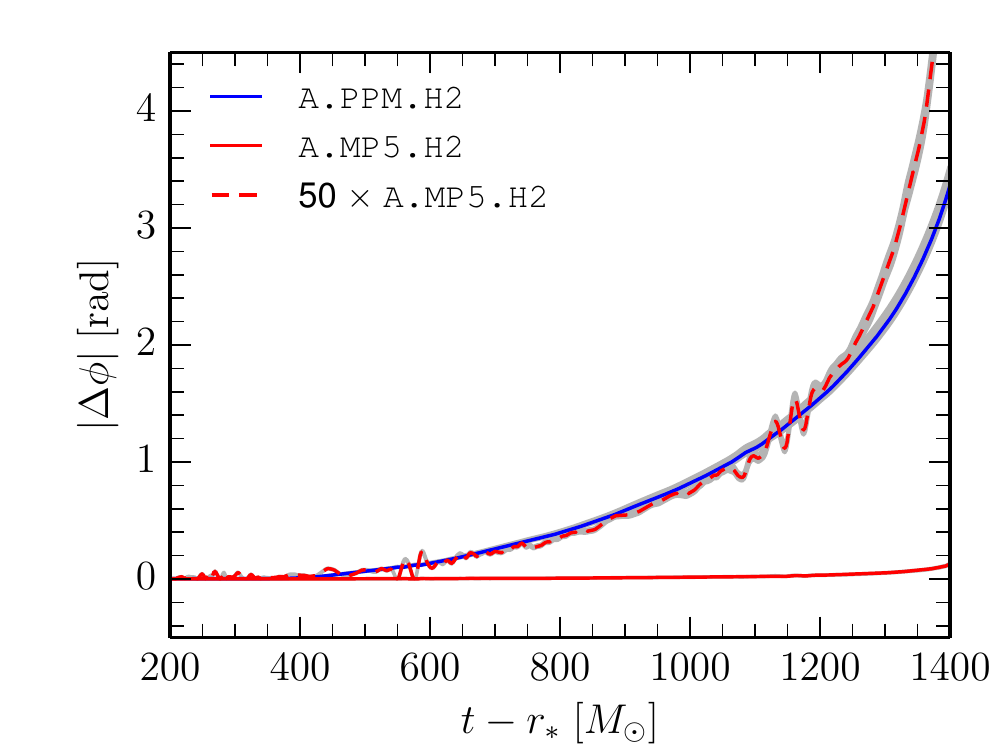}
  \caption{\label{fig:bns45.thc.vs.whisky} Estimated phase uncertainty
    due to finite resolution for \texttt{Whisky} and \texttt{WhiskyTHC}
    in the case of binary A at the common resolution of $h =
    0.2\ M_\odot$. The blue line shows the phase differences between
    \texttt{Whisky} and its Richardson extrapolated waveforms computed
    assuming convergence order between 1.7. The red line shows the phase
    differences between \texttt{WhiskyTHC} and its Richardson
    extrapolated waveforms computed assuming convergence order between
    2.8. The dashed red line shows the same data as the red one, but
    rescaled by a factor $50$. Finally, the grey-shaded regions show the
    estimated uncertainty, computed by varying the order used for the
    Richardson extrapolation by $\pm 0.2$.}
\end{center}
\end{figure}

As a consequence of having a higher convergence order, \texttt{WhiskyTHC}
is also significantly more accurate. This is shown in Figure
\ref{fig:bns45.thc.vs.whisky}, where we compare the estimated phase error
between the runs \texttt{A.MP5.H2}, \texttt{A.PPM.H2}, and the Richardson
extrapolated phase from \texttt{WhiskyTHC} and \texttt{Whisky},
respectively. We roughly estimate the uncertainty in this procedure by
performing two different extrapolations for each code, varying the
convergence order by $\pm 0.2$ with respect to the estimated one. The
resulting range of phase errors are shown as shaded regions in the
figure. Clearly, the simulation carried out with \texttt{Whisky} has an
uncertainty in phase which is almost equivalent to one gravitational-wave
cycle, \ie of the order of $\sim 7\ \%$ of the entire accumulated phase
(\cf line \texttt{A.PPM.H2}). At the same time, \texttt{WhiskyTHC} has an
error which, at the same resolution and for comparable computational
costs, is $\sim 50$ times smaller than \texttt{Whisky} (\cf
\texttt{A.MP5.H2}). We should note that we have also tried to estimate
the phase error for \texttt{WhiskyTHC} using the Richardson-extrapolated
data obtained with the \texttt{Whisky} code and assuming convergence
order of $1.7$. In this case, we found an even smaller estimated phase
error, but with an uncertainty, measured by varying the order in the
extrapolation by $\pm 0.1$, of more than $100\ \%$. As a final remark, we
note that our error estimates only reflect the numerical truncation
error. Other systematic errors and, in particular, finite extraction
radius effects, inaccuracies in the initial data (\eg eccentricity) are
also present and might be relevant, especially for \texttt{WhiskyTHC}. On
the other hand, because we are here interested only in evaluating and
comparing the accuracy of the two numerical methods, we can ignore these
systematic contributions.

\begin{figure}
\begin{center}
  \includegraphics[width=0.7\textwidth]{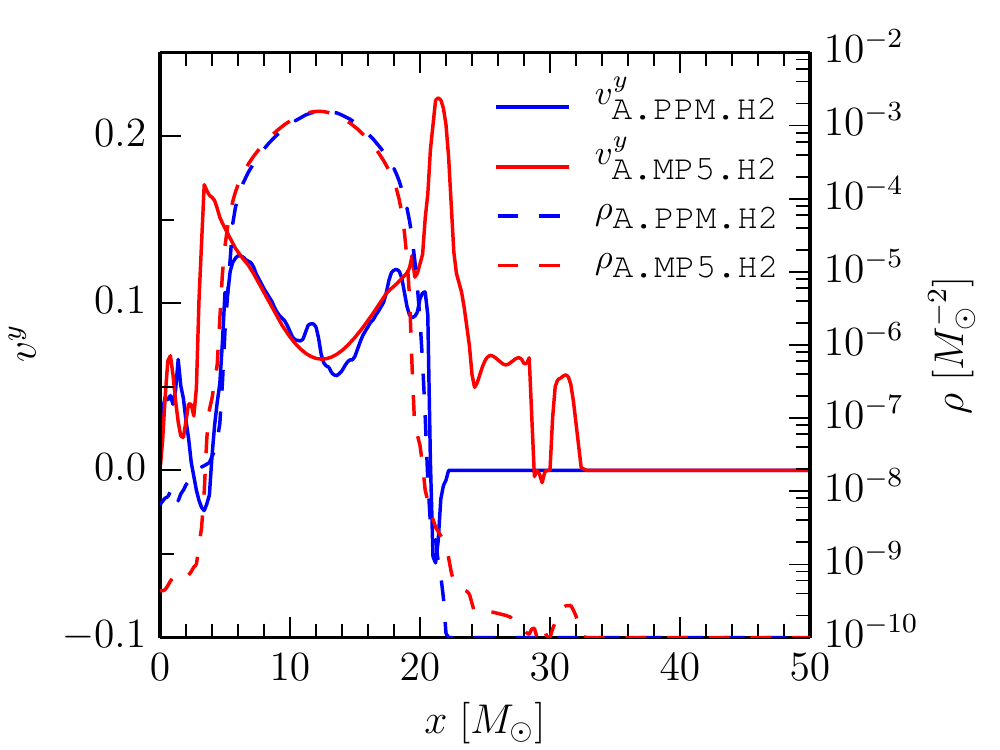}
  \caption{\label{fig:bns45.profile} One-dimensional cuts along the $x$
    axis of rest-mass density (dashed) and $y$ component of the velocity,
    $v^y$, (solid) lines for \texttt{Whisky} (blue) and
    \texttt{WhiskyTHC} (red) for the binary A. The velocity scale is
    shown on the left, while the rest-mass density scale is shown on the
    right. The data is taken after $\simeq 1$ orbit, at the approximate
    moment when the center of one of the two stars crosses the $x$
    axis. Notice that we correct for the de-phasing accumulated between
    \texttt{Whisky} and \texttt{WhiskyTHC} by taking the data at slightly
    different times: $t = 625.92\ M_\odot$ and $t = 622.08\ M_\odot$
    respectively.}
\end{center}
\end{figure}

A physical insight on why the \texttt{Whisky} code has a considerably
lower accuracy than the \texttt{WhiskyTHC} code can be gained by looking
at Figure \ref{fig:bns45.profile}. There, we show one-dimensional cuts of
the rest-mass density (solid lines) and of the $y$ component of the
velocity, $v^y$, (dashed lines) for runs \texttt{A.MP5.H2} and
\texttt{A.PPM.H2} along the $x$ axis. The data is taken at the
approximate time when the first orbit is completed and the centers of the
two stars are aligned along the axis. Since this happens at two different
times for \texttt{A.MP5.H2} (blue lines) and \texttt{A.PPM.H2} (red
lines) , the lines in the figure show data taken at two slightly
different coordinate times. Particularly important in this comparison is
the velocity profile inside the stars, as the velocity in the low-density
atmosphere around the stars is in any case dominated by unphysical
numerical effects as neither of the two codes is able to provide a
reasonable description of the stellar surface. The velocity in the
interior, on the other hand, is directly related to the orbital motion of
the two stars and hence to the gravitational-wave phase.

As can be seen from the figure, the \texttt{Whisky} code is not able to
transport the velocity profile of the star correctly; rather, the higher
numerical dissipation has the effect of slowly ``flattening'' both the
rest-mass density and the linear-momentum profiles in the stellar cores.
Since this flattening proceeds at different rates for the two different
fields, it results in a distortion of the linear velocity profile (and of
other related physical quantities), which, in turn, results in a small
deformation of the stars. In this case, the artificial deformation is
such that the stars become less compact, thus less efficient in losing
angular momentum via gravitational waves and hence delaying the time of
merger. We expect this deformation to be the leading source of error in
the phase evolution for the \texttt{Whisky} code, at least for the
high-compactness stars considered here.

\section{Conclusions}
\label{sec:grthc.conclusions}

We have presented a new multi-dimensional, general-relativistic
hydrodynamic code, \texttt{WhiskyTHC}, born from the merger of the
\texttt{Whisky} and the \texttt{THC} codes. This code inherited from
\texttt{Whisky} the primitive recovery routine as well as a new EOS
framework with support for composition and energy dependent realistic
equation of state \cite{Galeazzi2013}, and from \texttt{THC} the use of
high-order flux-vector splitting finite-differencing schemes
\cite{Radice2012a}. This is the first genuinely higher-than-second-order
fully general-relativistic code.

Amongst the new techniques introduced with \texttt{WhiskyTHC} is the use
of positivity-preserving limiters \cite{Hu2013} as a way to treat
low-density regions alternative to the traditional ``atmosphere''
prescriptions. We have shown that this treatment is able to significantly
improve the quality of simulations involving isolated neutron stars and
to effectively remove the loss of the conservative properties of the
equations in the boundary between the stellar surface and the ambient
atmosphere. Because the introduction of positivity preserving limiters in
any hydrodynamics code is rather straightforward and allows one to use
arbitrarily small values of the atmosphere, we recommend its use as
the method of choice for the evolution of compact stars.

We have also demonstrated the accuracy of our code in a series of
classical tests involving the linear and nonlinear evolution of isolated
neutron stars. In particular, we have shown that our code is able to
stably evolve isolated stars for a long time and can attain high order
(\ie third) of convergence in the simulation of the gravitational
collapse of nonrotating stars to black holes.

Finally, we have applied our code to the simulation of the late-inspiral
and merger of two neutron stars in quasi-circular orbits. Using rather
small-separation binaries to test the dependence of the results on the
atmosphere treatment, we have shown that our results are completely
independent of it. Furthermore, we have demonstrated the higher order of
convergence and accuracy of our new code when compared with our old
\texttt{Whisky} code, which implements the standard second-order schemes
that are commonly employed for matter simulations in numerical
relativity. In particular, we have found higher than second-order
convergence in the phase and an overall phase error which, at the same
resolution and with similar computational costs, is estimated to be
$\simeq 50$ times smaller than the one for \texttt{Whisky} for the
binaries we considered and at moderate resolution.

Our future plans include the exploitation of the efficiency of the
high-order methods in \texttt{WhiskyTHC} to pursue a systematic
investigation of tidal effects in binary-neutron-stars mergers, as well
as in black-hole neutron star binaries, using realistic equation of
states and compactness parameters. We also plan to carefully assess the
detectability of such effects by advanced gravitational-wave detectors.

\ack 
It is a pleasure to acknowledge W.\ Kastaun for kindly providing the
primitive recovery routine, F. Pannarale for providing the PN waveforms.
We also thank I.\ Hawke, D.\ Alic and K.\ Takami for numerous useful
discussions. Partial support comes from a Sherman Fairchild Foundation
grant to Caltech, the VESF grant (EGO-DIR-69-2010), the DFG grant
SFB/Transregio 7, and by the CompStar network, COST Action MP1304. The
calculations were performed on the SuperMUC cluster at the LRZ, on the
Datura cluster at the AEI, and on the LOEWE cluster in Frankfurt.

\section*{References}
\providecommand{\newblock}{}


\begin{thebibliography}{10}
\expandafter\ifx\csname url\endcsname\relax
  \def\url#1{{\tt #1}}\fi
\expandafter\ifx\csname urlprefix\endcsname\relax\def\urlprefix{URL }\fi
\providecommand{\eprint}[2][]{\url{#2}}

\bibitem{Radice2013b}
{Radice} D, {Rezzolla} L and {Galeazzi} F 2014 {\em Mon. Not. R. Astron. Soc.
  L.\/} {\bf 437} L46--L50 (\textit{Preprint} \eprint{1306.6052})

\bibitem{Baiotti:2010}
Baiotti L, Damour T, Giacomazzo B, Nagar A and Rezzolla L 2010 {\em Phys. Rev.
  Lett.\/} {\bf 105} 261101 (\textit{Preprint} \eprint{1009.0521})

\bibitem{Baiotti2011}
{Baiotti} L, {Damour} T, {Giacomazzo} B, {Nagar} A and {Rezzolla} L 2011 {\em
  Phys. Rev. D\/} {\bf 84} 024017 (\textit{Preprint} \eprint{1103.3874})

\bibitem{Bernuzzi2012}
{Bernuzzi} S, {Nagar} A, {Thierfelder} M and {Br{\"u}gmann} B 2012 {\em Phys.
  Rev. D\/} {\bf 86} 044030 (\textit{Preprint} \eprint{1205.3403})

\bibitem{Hotokezaka2013b}
{Hotokezaka} K, {Kyutoku} K and {Shibata} M 2013 {\em Phys. Rev. D\/} {\bf 87}
  044001 (\textit{Preprint} \eprint{1301.3555})

\bibitem{Aylott:2009ya}
Aylott B {\em et~al.\/} 2009 {\em Class. Quantum Grav.\/} {\bf 26} 165008
  (\textit{Preprint} \eprint{0901.4399})

\bibitem{Hinder2013}
{Hinder} I and et~al 2013 {\em Classical and Quantum Gravity\/} {\bf 31} 025012
  (\textit{Preprint} \eprint{1307.5307})

\bibitem{Radice2012a}
{Radice} D and {Rezzolla} L 2012 {\em Astron. Astrophys.\/} {\bf 547} A26
  (\textit{Preprint} \eprint{1206.6502})

\bibitem{Tchekhovskoy2007}
Tchekhovskoy A, McKinney J~C and Narayan R 2007 {\em Mon. Not. R. Astron.
  Soc.\/} {\bf 379} 469--497

\bibitem{DelZanna2007}
{Del Zanna} L, {Zanotti} O, {Bucciantini} N and {Londrillo} P 2007 {\em Astron.
  Astrophys.\/} {\bf 473} 11--30 (\textit{Preprint} \eprint{0704.3206})

\bibitem{Bucciantini2011}
{Bucciantini} N and {Del Zanna} L 2011 {\em Astron. Astrophys.\/} {\bf 528}
  A101 (\textit{Preprint} \eprint{1010.3532})

\bibitem{MTW1973}
Misner C~W, Thorne K~S and Wheeler J~A 1973 {\em Gravitation\/} (San Francisco:
  W. H. Freeman)

\bibitem{Rezzolla_book:2013}
{Rezzolla} L and {Zanotti} O 2013 {\em {Relativistic Hydrodynamics}\/} (Oxford
  University Press, Oxford UK)

\bibitem{Bona:2010is}
Bona C, Bona-Casas C and Palenzuela C 2010 {\em Phys. Rev. D\/} {\bf 82} 124010
  (\textit{Preprint} \eprint{1008.0747})

\bibitem{Bona-and-Palenzuela-Luque-2005:numrel-book}
Bona C and Palenzuela-Luque C 2005 {\em Elements of Numerical Relativity\/}
  (Berlin: Springer-Verlag)

\bibitem{Gundlach2005:constraint-damping}
Gundlach C, Martin-Garcia J~M, Calabrese G and Hinder I 2005 {\em Classical
  Quantum Gravity\/} {\bf 22} 3767--3774 (\textit{Preprint}
  \eprint{gr-qc/0504114})

\bibitem{Alic:2011a}
{Alic} D, {Bona-Casas} C, {Bona} C, {Rezzolla} L and {Palenzuela} C 2012 {\em
  Phys. Rev. D\/} {\bf 85} 064040 (\textit{Preprint} \eprint{1106.2254})

\bibitem{Loffler:2011ay}
L{\"{o}}ffler F, Faber J, Bentivegna E, Bode T, Diener P, Haas R, Hinder I,
  Mundim B~C, Ott C~D, Schnetter E, Allen G, Campanelli M and Laguna P 2012
  {\em Class. Quantum Grav.\/} {\bf 29} 115001 (\textit{Preprint}
  \eprint{arXiv:1111.3344 [gr-qc]})

\bibitem{mclachlanweb41}
\url{http://www.cct.lsu.edu/~eschnett/McLachlan/index.html}

\bibitem{Bona95b}
{Bona} C, {Mass{\'o}} J, {Seidel} E and {Stela} J 1995 {\em Phys. Rev. Lett.\/}
  {\bf 75} 600--603 (\textit{Preprint} \eprint{gr-qc/9412071})

\bibitem{vanMeter:2006vi}
van Meter J~R, Baker J~G, Koppitz M and Choi D~I 2006 {\em Phys. Rev. D\/} {\bf
  73} 124011 (\textit{Preprint} \eprint{gr-qc/0605030})

\bibitem{Galeazzi2013}
{Galeazzi} F, {Kastaun} W, {Rezzolla} L and {Font} J~A 2013 {\em Phys. Rev.
  D\/} {\bf 88} 064009 (\textit{Preprint} \eprint{1306.4953})

\bibitem{Banyuls97}
Banyuls F, Font J~A, Ib{\'a}{\~n}ez J~M, Mart{\'\i} J~M and Miralles J~A 1997
  {\em Astrophys. J.\/} {\bf 476} 221

\bibitem{Baiotti04}
Baiotti L, Hawke I, Montero P~J, L{\"o}ffler F, Rezzolla L, Stergioulas N, Font
  J~A and Seidel E 2005 {\em Phys. Rev. D\/} {\bf 71} 024035

\bibitem{Kreiss73}
Kreiss H~O and Oliger J 1973 {\em Methods for the approximate solution of time
  dependent problems\/} (Geneva: GARP publication series No. 10)

\bibitem{Brown:2008sb}
Brown D, Diener P, Sarbach O, Schnetter E and Tiglio M 2009 {\em Phys. Rev.
  D\/} {\bf 79} 044023 (\textit{Preprint} \eprint{0809.3533})

\bibitem{Shu88}
Shu C~W and Osher S~J 1988 {\em J. Comput. Phys.\/} {\bf 77} 439

\bibitem{Shu97}
Shu C~W 1997 {E}ssentially non-oscillatory and weighted essentially
  non-oscillatory schemes for hyperbolic conservation laws Lecture notes ICASE
  Report 97-65; NASA CR-97-206253 NASA Langley Research Center

\bibitem{LeVeque92}
Leveque R~J 1992 {\em Numerical Methods for Conservation Laws\/} (Basel:
  Birkhauser Verlag)

\bibitem{Quirk1994}
Quirk J 1994 {\em International Journal for Numerical Methods in Fluids\/} {\bf
  18} 555--574

\bibitem{Anile_book}
{Anile} A~M 1990 {\em {Relativistic Fluids and Magneto-fluids}\/} (Cambridge
  University Press)

\bibitem{Font08}
Font J~A 2008 {\em Living Rev. Relativ.\/} {\bf 6} 4;
  http://www.livingreviews.org/lrr--2008--7 (\textit{Preprint}
  \eprint{0704.2608.})

\bibitem{Schnetter-etal-03b}
Schnetter E, Hawley S~H and Hawke I 2004 {\em Class. Quantum Grav.\/} {\bf 21}
  1465--1488

\bibitem{Goodale02a}
Goodale T, Allen G, Lanfermann G, Mass{\'o} J, Radke T, Seidel E and Shalf J
  2003 {\em Vector and Parallel Processing -- VECPAR'2002, 5th International
  Conference, Lecture Notes in Computer Science\/} (Berlin: Springer)

\bibitem{Berger84}
Berger M~J and Oliger J 1984 {\em J. Comput. Phys.\/} {\bf 53} 484--512

\bibitem{Berger89}
Berger M~J and Colella P 1989 {\em J. Comput. Phys.\/} {\bf 82} 64--84

\bibitem{Reisswig2012b}
{Reisswig} C, {Haas} R, {Ott} C~D, {Abdikamalov} E, {M{\"o}sta} P, {Pollney} D
  and {Schnetter} E 2013 {\em Phys. Rev. D\/} {\bf 87} 064023
  (\textit{Preprint} \eprint{1212.1191})

\bibitem{galeazzi_master}
Galeazzi F 2008 {\em Modelling fluid interfaces in numerical relativistic
  hydrodynamics\/} Master's thesis Universit{\`{a}} degli studi di Padova

\bibitem{kastaun_2006_hrs}
Kastaun W 2006 {\em Phys. Rev. D\/} {\bf 74} 124024

\bibitem{Millmore2010}
{Millmore} S~T and {Hawke} I 2010 {\em Classical Quantum Gravity\/} {\bf 27}
  015007 (\textit{Preprint} \eprint{0909.4217})

\bibitem{Font02c}
Font J~A, Goodale T, Iyer S, Miller M, Rezzolla L, Seidel E, Stergioulas N,
  Suen W~M and Tobias M 2002 {\em Phys. Rev. D\/} {\bf 65} 084024
  (\textit{Preprint} \eprint{gr-qc/0110047})

\bibitem{Radice2011}
{Radice} D and {Rezzolla} L 2011 {\em Phys. Rev. D\/} {\bf 84} 024010
  (\textit{Preprint} \eprint{1103.2426})

\bibitem{Hu2013}
Hu X~Y, Adams N~a and Shu C~W 2013 {\em Journal of Computational Physics\/}
  {\bf 242} 169--180

\bibitem{Zhang2010}
{Zhang} X and {Shu} C~W 2010 {\em Journal of Computational Physics\/} {\bf 229}
  8918--8934

\bibitem{Zhang2011}
Zhang X and Shu C~W 2011 {\em Proceedings of the Royal Society A: Mathematical,
  Physical and Engineering Sciences\/} {\bf 467} 2752--2776

\bibitem{Zhang2011a}
Zhang X and Shu C~W 2011 {\em Journal of Computational Physics\/} {\bf 230}
  1238--1248

\bibitem{Balsara2012}
{Balsara} D~S 2012 {\em Journal of Computational Physics\/} {\bf 231}
  7504--7517

\bibitem{Tolman39}
Tolman R~C 1939 {\em Phys. Rev.\/} {\bf 55} 364

\bibitem{Oppenheimer39b}
Oppenheimer J~R and Volkoff G 1939 {\em Phys. Rev.\/} {\bf 55} 374

\bibitem{Baiotti03a}
Baiotti L, Hawke I, Montero P and Rezzolla L 2003 {\em Computational
  Astrophysics in Italy: Methods and Tools\/} vol~1 ed Capuzzo-Dolcetta R
  (Trieste: MSAIt) p 210

\bibitem{Cordero2009}
{Cordero-Carri{\'o}n} I, {Cerd{\'a}-Dur{\'a}n} P, {Dimmelmeier} H, {Jaramillo}
  J~L, {Novak} J and {Gourgoulhon} E 2009 {\em Phys. Rev. D\/} {\bf 79} 024017
  (\textit{Preprint} \eprint{0809.2325})

\bibitem{Thierfelder2011}
Thierfelder M, Bernuzzi S and Br\"{u}gmann B 2011 {\em Phys. Rev. D\/} {\bf 84}
  1--30

\bibitem{liebling_2010_emr}
{Liebling} S~L, {Lehner} L, {Neilsen} D and {Palenzuela} C 2010 {\em Phys. Rev.
  D\/} {\bf 81} 124023 (\textit{Preprint} \eprint{1001.0575})

\bibitem{Radice:10}
{Radice} D, {Rezzolla} L and {Kellerman} T 2010 {\em Classical Quantum
  Gravity\/} {\bf 27} 235015 (\textit{Preprint} \eprint{1007.2809})

\bibitem{Kastaun2013}
{Kastaun} W, {Galeazzi} F, {Alic} D, {Rezzolla} L and {Font} J~A 2013 {\em
  Phys. Rev. D\/} {\bf 88} 021501 (\textit{Preprint} \eprint{1301.7348})

\bibitem{Baiotti06}
Baiotti L and Rezzolla L 2006 {\em Phys. Rev. Lett.\/} {\bf 97} 141101
  (\textit{Preprint} \eprint{gr-qc/0608113})

\bibitem{Baiotti07}
Baiotti L, Hawke I and Rezzolla L 2007 {\em Classical Quantum Gravity\/} {\bf
  24} S187--S206 (\textit{Preprint} \eprint{gr-qc/0701043})

\bibitem{Thierfelder10}
{Thierfelder} M, {Bernuzzi} S, {Hilditch} D, {Bruegmann} B and {Rezzolla} L
  2010 {\em Phys. Rev. D\/} {\bf 83} 064022 (\textit{Preprint}
  \eprint{1012.3703})

\bibitem{Alic2013}
{Alic} D, {Kastaun} W and {Rezzolla} L 2013 {\em Phys. Rev. D\/} {\bf 88}
  064049 (\textit{Preprint} \eprint{1307.7391})

\bibitem{Colella84}
Colella P and Woodward P~R 1984 {\em J. Comput. Phys.\/} {\bf 54} 174

\bibitem{Harten83}
Harten A, Lax P~D and van Leer B 1983 {\em SIAM Rev.\/} {\bf 25} 35

\bibitem{Einfeldt88}
Einfeldt B 1988 {\em SIAM J. Numer. Anal.\/} {\bf 25} 294--318

\bibitem{Baiotti:2010ka}
Baiotti L, Shibata M and Yamamoto T 2010 {\em Phys. Rev. D\/} {\bf 82} 064015
  (\textit{Preprint} \eprint{1007.1754})

\bibitem{Gourgoulhon01}
Gourgoulhon E, Grandcl{\'e}ment P, Taniguchi K, Marck J~A and Bonazzola S 2001
  {\em Phys. Rev. D\/} {\bf 63} 064029

\bibitem{lorene41}
\url{http://www.lorene.obspm.fr}

\bibitem{Baiotti08}
{Baiotti} L, {Giacomazzo} B and {Rezzolla} L 2008 {\em Phys. Rev. D\/} {\bf 78}
  084033 (\textit{Preprint} \eprint{0804.0594})

\bibitem{Boyle:2009vi}
{Boyle} M and {Mrou{\'e}} A~H 2009 {\em Phys. Rev. D\/} {\bf 80} 124045
  (\textit{Preprint} \eprint{0905.3177})

\bibitem{Reisswig:2009us}
Reisswig C, Bishop N~T, Pollney D and Szilagyi B 2009 {\em Phys. Rev. Lett.\/}
  {\bf 103} 221101 (\textit{Preprint} \eprint{0907.2637})

\bibitem{Reisswig:2010a}
Reisswig C, Bishop N~T, Pollney D and Szilagyi B 2010 {\em Class. Quantum
  Grav.\/} {\bf 27} 075014 (\textit{Preprint} \eprint{arXiv:0912.1285})

\bibitem{Reisswig:2011}
Reisswig C and D P 2011 {\em Class. Quantum Grav.\/} {\bf 28} 195015
  (\textit{Preprint} \eprint{1006.1632})

\bibitem{Kyutoku2012}
Kyutoku K, Ioka K and Shibata M 2013 {\em Mon.~Not.~R.~Astron.~Soc.~L.\/} {\bf
  437} L6--L10

\bibitem{Baiotti:2009gk}
Baiotti L, Giacomazzo B and Rezzolla L 2009 {\em Class. Quantum Grav.\/} {\bf
  26} 114005 (\textit{Preprint} \eprint{0901.4955})

\end{thebibliography}
\end{document}